\documentclass[journal]{IEEEtran}
\usepackage{bbm}
\usepackage{cite}
\usepackage{amsfonts}
\usepackage{mathrsfs}
\usepackage{graphicx}
\usepackage{amssymb}
\usepackage{latexsym}
\usepackage{amsmath}
\usepackage{stfloats}
\usepackage{cases}
\usepackage{setspace}
\usepackage{bigstrut}
\usepackage{bm}
\usepackage{url}
\usepackage{color}
\usepackage{subfigure}
\usepackage[ruled,linesnumbered]{algorithm2e}
\usepackage{multirow}
\usepackage{amsthm}

\newcommand{\bl}[1]{\color{blue}#1}

\begin{document}

\title{Wideband Precoding for U6G XL-MIMO Systems: Beam Squint Boundaries and Channel Slicing}
    \author{Zhizheng Lu,~\IEEEmembership{Student Member, IEEE}, Yu Han,~\IEEEmembership{Member, IEEE}, Xiaojie Li,~\IEEEmembership{Student Member, IEEE},
    
    Shi Jin,~\IEEEmembership{Fellow, IEEE}, and Michail Matthaiou,~\IEEEmembership{Fellow, IEEE}
    \thanks{Z. Lu, Y. Han, X. Li, and S. Jin are with the School of Information Science and Engineering, Southeast University, Nanjing 210096, China (e-mail: luzz@seu.edu.cn; hanyu@seu.edu.cn; xiaojieli@seu.edu.cn; jinshi@seu.edu.cn).}
    \thanks{M. Matthaiou is with the Centre for Wireless Innovation (CWI), Queen’s University Belfast, BT3 9DT, Belfast, U.K. (e-mail: m.matthaiou@qub.ac.uk).}}
\maketitle

\begin{abstract}

The unconventionally large aperture of extremely large-scale multiple-input multiple-output (XL-MIMO) arrays, in conjunction with the wider bandwidths in the upper-6 GHz (U6G) frequency bands, will very likely lead to non-negligible beam squint effects. In the context of a limited number of radio frequency (RF) chains, and by adopting hybrid precoding, the beams at different subcarriers may point to different positions and compromise the spectral efficiency (SE). Moreover, the existence of \textit{multiple paths} in U6G XL-MIMO channels also entails practical challenges for \textit{wideband precoding}. It is therefore essential to ascertain whether the beam squint effect is pronounced for U6G XL-MIMO systems and design efficient wideband precoding schemes. To address these challenges, precise antenna-domain and frequency-domain \textit{wideband boundaries} are derived from the near-field and far-field perspectives, respectively. These boundaries can inform the design of wideband precoding in future system settings. Subsequently, a \textit{channel slicing} scheme is proposed for wideband precoding. The process involves the segmentation of U6G XL-MIMO channels into multiple blocks, with the objective of mitigating the beam squint effect for each path. The antenna-domain and frequency-domain slicing methods are developed for multipath and multiuser scenarios, respectively. The simulation results prove that the beam squint effect remains a significant issue for U6G XL-MIMO systems, while the near-field effect invariably precedes the beam squint effect as the array size and the bandwidth increase. In addition, the proposed scheme can greatly improve the SE.

\end{abstract}

\section{Introduction}\label{Sec: Introduction}

To underpin the significantly higher performance demands of the sixth generation (6G) wireless systems, which are expected to achieve a 100-fold increase in peak data rates (reaching the Tb/s level), and a tenfold reduction in latency \cite{6G}, the extremely large-scale multiple-input multiple-output (XL-MIMO) concept is perceived as a promising technology \cite{XL-MIMO1,XL-MIMO2,XL-MIMO3}. Through employing a greatly larger number of antennas at base stations (BSs), the spectral efficiency (SE), capacity, and coverage can be significantly improved \cite{SE1,SE2,SE3}. Meanwhile, the higher frequency bands, such as the upper-6 GHz (U6G) frequency band \cite{mid-band}, is also regarded as a key technology for 6G, which can provide richer spectrum resources compared to the fifth generation (5G) wireless systems, to further enhance the communication quality \cite{U6G1,U6G2}. It is important to note that it has a much reduced path loss compared to the millimeter wave (mmWave) and terahertz (THz) frequency bands, thus supporting greater coverage for 6G BSs. We recall that recent developments from the World Radiocommunication Conference 2023 (WRC-23) \cite{U6G3} are notable, with the allocation of up to 700 MHz of spectrum in the U6G frequency band (6.425-7.125 GHz) for exclusive and/or shared use, which is anticipated to be the primary frequency for IMT services in the 6G era. Based on the above discussion, it becomes indispensable to understand and eventually optimize key communication technologies for U6G XL-MIMO systems.

It is important to note that due to the extremely large number of antennas, fully digital precoding architectures will involve a high hardware cost. To address this critical issue, low-cost hybrid precoding architectures, with a small number of radio-frequency (RF) chains, are considered for XL-MIMO systems \cite{hybrid_architecture1,hybrid_architecture2}. In narrowband XL-MIMO systems,\footnote{Note that in this paper, a narrowband system does not necessarily point to a single subcarrier, but instead to systems where the beam squint effect across the entire array and bandwidth is negligible. In contrast, a wideband system refers to that the beam squint effect is nonnegligible.} frequency-independent analog precoders (which are constructed by analog phase shifters) can generate directional beams aligned with the physical directions/positions of the channel path components, to obtain the maximum SE. However, in the higher frequency bands and with the larger bandwidths considered for 6G systems, the beam squint effect will be amplified \cite{beamsquint1,beamsquint2, beamsquint3,beamsquint4,beamsquint6}, which means that the beams at different subcarriers will point to different physical directions/positions due to the frequency-independent analog precoders. Hence, when analog beams are designed to point at a desired user equipment (UE) under beam squint, only the signals around the central frequency will be effectively captured by the receiver. This will inevitably result in considerable degradation in the achievable SE.

Recently, many studies have been reported on the beam squint effect in the mmWave and THz systems \cite{work1,work2,lens_far,TTD_far0,TTD_far1,TTD_far2,TTD_far3,TTD_frequency_slicing,multiple_beam,TTD_hybrid,TTD_decouple,no_TTD,beamsquint_lens,beamsquint_bound,compared_method}, in which massive MIMO/XL-MIMO with hybrid precoding architectures were considered. In \cite{work1}, the wideband beamforming vectors were designed by maximizing the minimum array gain achieved in the entire bandwidth, while \cite{work2} utilized a semidefinite relaxation method to maximize the total array gain across all subcarriers. The authors in \cite{lens_far} proposed a novel transceiver design based on lens antenna subarrays and analog sub-band filters to compensate for the beam squint effect. In \cite{TTD_far0,TTD_far1,TTD_far2,TTD_far3,TTD_frequency_slicing,multiple_beam,TTD_hybrid,TTD_decouple}, true-time-delay (TTD) lines were employed to mitigate the beam squint effect in wideband systems. In \cite{TTD_far0} and \cite{TTD_far1}, the wideband hybrid beamforming approaches, based on virtual subarray and TTD lines, were proposed to eliminate the beam squint in the THz frequency band. The authors in \cite{TTD_far2} conceived a TTD-based THz hardware structure, which can eliminate the array gain loss caused by beam squint. The authors in \cite{TTD_far3} introduced a low-complexity beam squint mitigation scheme based on TTD lines, and proposed a wideband channel estimation algorithm with low training overhead. The authors in \cite{TTD_frequency_slicing} and \cite{multiple_beam} indicated that the deployment of TTD lines can facilitate the use of beam squint to serve multiple UEs simultaneously. In \cite{TTD_hybrid}, a beam steering approach was proposed based on the combination of TTD lines and phase shifters to enable continuous beam steering with minimal beam squint. In \cite{TTD_decouple}, a TTD-incorporated analog beamfocusing technique was designed, which can address the interplay between near-field propagation and wideband beam squint. However, the authors in \cite{no_TTD} indicated that although the TTD-lines-based architectures can provide higher SE, they will also consume more energy and exhibit poor energy efficiency. In \cite{beamsquint_lens}, a RF lens-based architecture was proposed, which was shown to be free from the beam squint effect. Moreover, the authors in \cite{beamsquint_bound} proposed a frequency-selective metric to redefine the far-field boundary in wideband systems. However, the solution was obtained based on a specific hardware architecture and beamforming for a single path, which cannot be directly utilized for U6G XL-MIMO systems with multiple paths. In \cite{compared_method}, a low-complexity spatial coding technique was proposed to mitigate the effect of beam squint and design the beamforming vector for near-field single-path channels.

The majority of existing works have focused on beam squint in the mmWave and THz frequency bands that involve only a single path. However, in the context of U6G XL-MIMO systems, it is imperative to determine whether the beam squint effect constitutes a significant concern within the U6G frequency bands. Moreover, it is also important to consider the presence of \textbf{\textit{multiple paths}} and \textbf{\textit{hybrid field effect}} when designing \textbf{\textit{wideband precoding}} for U6G XL-MIMO systems. Consequently, the precise boundaries between wideband and narrowband systems require urgent attention, while efficient design of wideband precoding is also a subject that requires attention. To the best knowledge of the authors, there has been no practical research progress on it.

In this study, a U6G XL-MIMO system with a hybrid precoding architecture is considered. The frequency-domain and antenna-domain \textbf{\textit{wideband boundaries}} in near-field and far-field perspectives are derived, and a multiple-domain \textbf{\textit{channel slicing}} scheme is then conceived to mitigate the beam squint effect of multiple paths and inform the design of wideband precoding. Compared with other works, the major contributions of this paper are summarized as follows:

\begin{itemize}
\item \textbf{\textit{Wideband boundaries:}} To evaluate whether the beam squint is worthy of attention for U6G XL-MIMO systems, the wideband boundaries, that can distinguish between wideband and narrowband systems, are theoretically derived from both the antenna-domain and frequency-domain perspectives. The multiple paths inherent in U6G XL-MIMO channels necessitates a focus on wideband boundaries from both near-field and far-field perspectives. The solutions can inform the design of efficient precoder for hybrid-field multipath channels in U6G XL-MIMO systems with hybrid precoding architectures.
\end{itemize}

\begin{itemize}
\item \textbf{\textit{Channel slicing:}} Based on the precise wideband boundaries, a channel slicing scheme is conceived for U6G XL-MIMO systems from the antenna-domain and frequency-domain perspectives, respectively. From the antenna-domain perspective, the constraint on subarrays is introduced to ensure a strong spatial resolution and mitigate the beam squint effect, while the constraint on bandwidths is also given from the frequency-domain perspective to mitigate the effects of beam squint and phase variations among different paths at different subcarriers.
\end{itemize}

\begin{itemize}
\item \textbf{\textit{Wideband precoding:}} Focusing on the inherent effect of multiple paths in U6G XL-MIMO channels, the antenna-domain and frequency-domain wideband precoding methods are proposed based on the channel slicing scheme, for scenarios where the scale of subarrays is adjustable and constant, respectively. The simulation results prove that the proposed schemes can improve the SE across the entire bandwidth, while achieving a higher SE for U6G XL-MIMO systems with multiple paths.
\end{itemize}

The rest of this paper is organized as follows: In Section $\rm \uppercase\expandafter{\romannumeral2}$, the channel model is introduced. In Section $\rm\uppercase\expandafter{\romannumeral3}$, precise antenna-domain and frequency-domain wideband boundaries are derived from near-field and far-field perspectives, respectively. In Section $\rm \uppercase \expandafter{\romannumeral4}$, a channel slicing scheme is proposed. Based on antenna- and frequency-domain slicing, efficient hybrid precoding schemes are designed. Section $\rm \uppercase \expandafter{\romannumeral5}$ evaluates the proposed wideband boundaries and channel slicing scheme. Finally, conclusions are given in Section $\rm \uppercase \expandafter{\romannumeral6}$.

\textit{Notations:} Henceforth, lower-case and upper-case bold letters denote vectors and matrices, respectively; $\otimes$ and $\odot$ represent the Kronecker product and the Hadamard product; ${{\bf{1}}_N}$ denotes the $N$-dimensional all-one vector. For a matrix ${\bf{A}}$, ${{\bf{A}}^{\rm{T}}}$ and ${{\bf{A}}^{\rm{H}}}$ represent the transpose and conjugate transpose operations; $\left[\bf A\right]_{n,:}$, $\left[\bf A\right]_{:,m}$, and $\left[\bf A\right]_{n,m}$ are the $n$-th row, $m$-th column, and $\left(n,m\right)$-th element of $\bf A$, while $\left[{\bf a}\right]_n$ is the $n$-th element of a vector ${\bf a}$. In addition, $\angle {\cdot} $ denotes taking the phase; $\left\lfloor \cdot \right\rfloor$ indicates the operation of rounding down; $\left| \cdot \right|$ and $\left\| \cdot \right\|$ represent taking the absolute value and the vector ${l_2}$ norm; ${\rm diag}\left({a}_1,\ldots,{a}_N\right)$ and ${\rm blkdiag}\left({\bf a}_1,\ldots,{\bf a}_N\right)$ are the diagonal and block diagonal matrices constructed by $\left\{{a}_1,\ldots,{a}_N\right\}$ and $\left\{{\bf a}_1,\ldots,{\bf a}_N\right\}$, respectively.

\section{Channel Model of U6G XL-MIMO Systems}\label{Sec: System Model}

\begin{figure}
  \centering
  \includegraphics[scale=0.4]{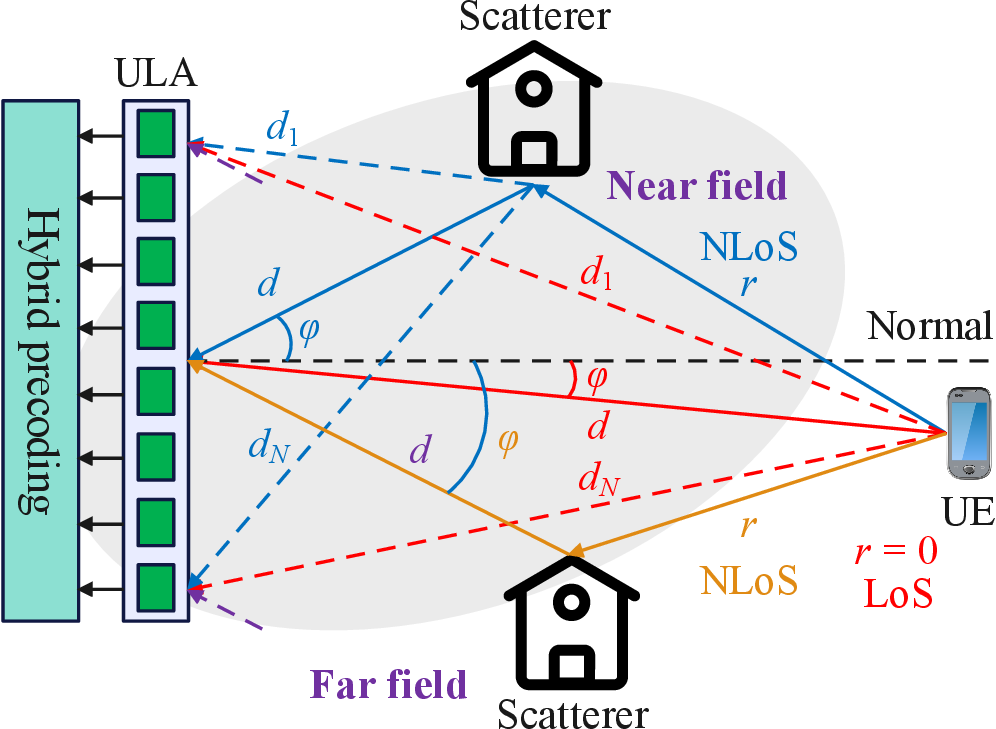}
  \caption{The considered U6G XL-MIMO channel model.} \label{Fig: channel model}
  \vspace{-0.4cm}
\end{figure}

In this paper, a U6G wideband XL-MIMO system with a hybrid precoding architecture is considered. The frequency of the central carrier is denoted as $f_{\rm c}$, and the wavelength is $\lambda _{\rm c} = \frac{c}{{{f_{\rm c}}}}$, where $c$ denotes the speed of light. The number of subcarriers is indicated by $M$, while the frequency spacing between adjacent subcarriers is $\Delta f = \frac{B}{M}$, where $B$ represents the bandwidth. The frequency of the $m$-th subcarrier can be denoted as $f_m = f_{\rm c} + \delta_{M,m}\Delta f$, where $\delta_{M,m} = m-\frac{M+1}{2}$, for $m=1,\ldots,M$. As shown in Fig.~\ref{Fig: channel model}, the BS is equipped with an $N$-antenna uniform linear array (ULA), in which the spacing between adjacent antennas is $s = \frac{\lambda_{\rm c}}{2}$. Without loss of generality, only a single-antenna UE is considered in this section, while the multiple-UE scenario and the specific hybrid precoding architecture will be introduced in Section $\rm \uppercase\expandafter{\romannumeral4}$.

In order to ascertain whether the beam squint effect remains pronounced in U6G XL-MIMO systems, a realistic transmission model for the channel is considered \cite{hybrid_architecture2}. Specifically, at the $m$-th subcarrier, the channel between the UE and the $n$-th antenna of the BS can be expressed as
\begin{equation}\label{Eq: Section2_1}
\begin{aligned}
&{\left[ {{{\bf{H}}_{{\rm{wn}}}}} \right]_{n,m}} = \sum\limits_{l = 1}^L {{\rho_l} {e^{j\frac{{2\pi }}{c}\left( {{f_{\rm{c}}} + {\delta _{M,m}}\Delta f} \right)\left( {{r_l} + {d_l} + \Delta {d_{n,l}}} \right)}}} \\ = & \sum\limits_{l=1}^L {{\rho_l}{e^{j\frac{{2\pi }}{c}\left( {{f_{\rm{c}}}\left( {{r_l} + {d_l}} \right) + {f_{\rm{c}}}\Delta {d_{n,l}} + {\delta _{M,m}}\Delta f\left( {{r_l} + {d_l}} \right) + {\delta _{M,m}}\Delta f\Delta {d_{n,l}}} \right)}}},
\end{aligned}
\end{equation}
for $n = 1,\ldots,N$, and $m = 1,\ldots,M$, where ${\bf H}_{\rm wn} \in \mathbb{C}^{N \times M}$ denotes the realistic wideband channel between the UE and the BS; $L$ is the number of multiple paths in the channel, containing a line-of-sight (LoS) path and $L-1$ non-LoS (NLoS) paths; ${\rho_l} \sim {\cal C}{\cal N}\left( {0,1} \right)$ denotes the path loss of the $l$-th path; $r_l \ge 0$ and $d_l > 0$ denote the range from the UE to the scatterer, and the distance from the scatterer to the center of the BS, respectively. Note that in the LoS path, $r_l = 0$, and the UE can also be considered as the scatterer of the LoS path. Moreover, $\Delta {d_{n,l}} = {d_{n,l}} - {d_l}$ represents the distance variation from the scatterer to the $n$-th antenna of the BS, and to the center of the ULA, where ${d_{n,l}}$ and ${d_l}$ indicate the distances from the scatterer to the $n$-th antenna of the BS, and to the center of the ULA, respectively. The distance from the scatterer to the $n$-th antenna of the BS can be calculated as
\begin{equation}\label{Eq: Section2_2}
{d_{n,l}} = \sqrt {d_l^2 - 2d_l{\delta _{N,n}}s{\theta_l} + \delta_{N,n}^2{s^2}},
\end{equation}
where $\delta_{N,n} = n - \frac{N+1}{2}$, for $n = 1,\ldots,N$; $\theta_l \in \left(-1,1\right)$ is the sine of the azimuth angle in the $l$-th path, such that $\theta_l = \sin \left(\varphi_l\right)$, where $\varphi_l \in \left( -\frac{\pi}{2}, \frac{\pi}{2} \right)$ is the azimuth angle between the path and the array normal. Note that the phase of ${\left[ {{{\bf{H}}_{{\rm{wn}}}}} \right]_{n,m}}$ can be divided into four parts, in which ${e^{j\frac{{2\pi }}{c}{f_{\rm{c}}}\left( {{r_l} + {d_l}} \right)}}$ is independent of the antenna index $n$ and the subcarrier index $m$. Thus, ${\left[ {{{\bf{H}}_{{\rm{wn}}}}} \right]_{n,m}}$ can be alternatively written as
\begin{equation}\label{Eq: Section2_3}
\begin{aligned}
&{\left[ {{{\bf{H}}_{{\rm{wn}}}}} \right]_{n,m}} \\= &\sum\limits_{l=1}^L {{g_l}{e^{j\frac{{2\pi }}{c}{f_{\rm{c}}}\Delta {d_{n,l}}}}{e^{j\frac{{2\pi }}{c}{f_{\rm{c}}}{\delta_{M,m}}\Delta f\left( {{r_l} + {d_l}} \right)}}{e^{j\frac{{2\pi }}{c}{\delta _{M,m}}\Delta f\Delta {d_{n,l}}}}},
\end{aligned}
\end{equation}
where the path complex gain is ${g_l} = {\rho_l}{e^{j\frac{{2\pi }}{c}{f_{\rm{c}}}\left( {{r_l} + {d_l}} \right)}}$. The first part of the phase in {\bl\eqref{Eq: Section2_3}}, i.e. $\frac{{2\pi }}{c}{f_{\rm{c}}}\Delta {d_{n,l}}$, characterizes the influence of distance variation from the scatterer to the center of the ULA and to the $n$-th antenna of the BS. This phenomenon is known as the near-field effect \cite{near-field/far-field}, which depends only on the antenna index $n$. The second part of the phase in {\bl\eqref{Eq: Section2_3}}, i.e. $\frac{{2\pi }}{c}{f_{\rm{c}}}{\delta _{M,m}}\Delta f\left( {{r_l} + {d_l}} \right)$, characterizes the influence of frequency variation between the central carrier and the $m$-th subcarrier \cite{multiple/single-carrier}, which depends only on the subcarrier index $m$. Furthermore, the third part of the phase in {\bl\eqref{Eq: Section2_3}}, i.e. $\frac{{2\pi }}{c}{\delta _{M,m}}\Delta f\Delta {d_{n,l}}$, is related to both the distance and frequency variations between different antennas and subcarriers, and can characterize the influence of the beam squint effect \cite{beamsquint1}. Subsequently, at the $m$-th subcarrier, the realistic channel model can be succinctly expressed as
\begin{equation}\label{Eq: Section2_4}
\begin{aligned}
&{\left[ {{{\bf{H}}_{{\rm{wn}}}}} \right]_{:,m}} \\= &\sum\limits_{l=1}^L {g_l} {e^{j\frac{{2\pi }}{c}{f_{\rm{c}}}{\delta_{M,m}}\Delta f\left( {{r_l} + {d_l}} \right)}}{\bf{w}}\left( {\theta_l,{d_l}} \right) \odot {{\bf{q}}_{\rm{c}}}\left( {m,{\theta_l},{d_l}} \right),
\end{aligned}
\end{equation}
where the antenna-domain spherical wave steering vector, ${\bf w}\left(\theta_l,d_l\right) \in \mathbb{C}^{N \times 1}$, is given by
\begin{equation}\label{Eq: Section2_5}
{\bf{w}}\left( {\theta_l ,d_l} \right) = {\left[ {{e^{j\frac{{2\pi }}{c}{f_{\rm{c}}}\Delta {d_{1,l}}}}, \cdots ,{e^{j\frac{{2\pi }}{c}{f_{\rm{c}}}\Delta {d_{N,l}}}}} \right]^{\rm{T}}},
\end{equation}
for $l = 1,\ldots,L$, while the beam squint column vector ${{\bf{q}}_{\rm{c}}}\left( {m,\theta_l ,d_l} \right) \in \mathbb{C}^{N \times 1}$ can be expressed as
\begin{equation}\label{Eq: Section2_6}
{{\bf{q}}_{\rm{c}}}\left( {m,\theta_l, d_l} \right) = {\left[ {{e^{j\frac{{2\pi }}{c}{\delta _{M,m}}\Delta f\Delta {d_{1,l}}}}, \cdots ,{e^{j\frac{{2\pi }}{c}{\delta _{M,m}}\Delta f\Delta {d_{N,l}}}}} \right]^{\rm{T}}},
\end{equation}
for $m = 1,\ldots,M$. Similarly, at the $n$-th antenna, the realistic channel model can be denoted by
\begin{equation}\label{Eq: Section2_7}
{\left[ {{{\bf{H}}_{{\rm{wn}}}}} \right]_{n,:}} = \sum\limits_{l=1}^L {{g_l}{e^{j\frac{{2\pi }}{c}{f_{\rm{c}}}\Delta {d_{n,l}}}}{{\bf{p}}^{\rm{T}}}\left( {{d_l},{r_l}} \right) \odot {\bf{q}}_{\rm{r}}^{\rm{T}}\left( {n,{\theta_l},{d_l}} \right)},
\end{equation}
where the delay-domain steering vector ${\bf p}\left(d_l,r_l\right) \in \mathbb{C}^{M \times 1}$ is
\begin{equation}\label{Eq: Section2_8}
{\bf{p}}\left( {d_l,r_l} \right) = {\left[ {{e^{j\frac{{2\pi }}{c}{\delta _{M,1}}\Delta f\left( {r_l + d_l} \right)}}, \cdots ,{e^{j\frac{{2\pi }}{c}{\delta _{M,M}}\Delta f\left( {r_l + d_l} \right)}}} \right]^{\rm{T}}},
\end{equation}
while the beam squint row vector ${{\bf{q}}_{\rm{r}}}\left( {n,\theta_l, d_l} \right) \in \mathbb{C}^{M \times 1}$ can be expressed according to
\begin{equation}\label{Eq: Section2_9}
{{\bf{q}}_{\rm{r}}}\left( {n,\theta_l ,d_l} \right) = {\left[ {{e^{j\frac{{2\pi }}{c}{\delta _{M,1}}\Delta f\Delta {d_{n,l}}}}, \cdots, {e^{j\frac{{2\pi }}{c}{\delta _{M,M}}\Delta f\Delta {d_{n,l}}}}} \right]^{\rm{T}}},
\end{equation}
for $n = 1,\ldots,N$. Consequently, the full-dimensional wideband channel can be modeled by
\begin{equation}\label{Eq: Section2_10}
{{\bf{H}}_{{\rm{wn}}}} = \sum\limits_{l=1}^L {{g_l}\left({\bf{w}}\left( {{\theta_l},{d_l}} \right){{\bf{p}}^{\rm{T}}}\left( {{d_l},{r_l}} \right)\right) \odot {{\bf{Q}}_{{\rm{wn}}}}\left( {\theta_l,d_l} \right)},
\end{equation}
where the beam squint matrix, ${\bf Q}_{\rm wn} \left( \theta, d \right) \in \mathbb{C}^{N \times M}$, is given by
\begin{equation}\label{Eq: Section2_11}
{{\bf{Q}}_{{\rm{wn}}}}\left( {\theta ,d} \right) = \left[ {\begin{array}{*{20}{c}}
{{e^{j\frac{{2\pi }}{c}{\delta _{M,1}}\Delta f\Delta {d_1}}}}& \cdots &{{e^{j\frac{{2\pi }}{c}{\delta _{M,M}}\Delta f\Delta {d_1}}}}\\
 \vdots & \ddots & \vdots \\
{{e^{j\frac{{2\pi }}{c}{\delta _{M,1}}\Delta f\Delta {d_N}}}}& \cdots &{{e^{j\frac{{2\pi }}{c}{\delta _{M,M}}\Delta f\Delta {d_N}}}}
\end{array}} \right].
\end{equation}
Alternatively, ${{\bf{Q}}_{{\rm{wn}}}}\left( {\theta ,d} \right) = \left[ {{{\bf{q}}_{\rm{c}}}\left( {1,\theta ,d} \right), \cdots ,{{\bf{q}}_{\rm{c}}}\left( {M,\theta ,d} \right)} \right]$ if it is observed from the frequency domain, and ${{\bf{Q}}_{{\rm{wn}}}}\left( {\theta ,d} \right) = \left[ {{{\bf{q}}_{\rm{r}}}\left( {1,\theta ,d} \right), \cdots ,{{\bf{q}}_{\rm{r}}}\left( {N,\theta ,d} \right)} \right]^{\rm T}$ if it is observed from the antenna domain, respectively. Note that ${\bf Q}_{\rm wn} \left( \theta, d \right)$ has modeled the specific influence of the beam squint effect on each path of the realistic channel model.

In conventional sub-6 GHz and massive MIMO systems, the near-field and beam squint effects as in {\bl\eqref{Eq: Section2_10}} are typically ignored in hybrid precoding design. However, the SE will be compromised in U6G XL-MIMO systems, as these effects may become amplified with a significant increase in bandwidth and number of antennas. Moreover, in contrast to the mmWave/THz systems, it is also imperative to consider the presence of multiple paths. In the following sections, precise \textbf{\textit{boundaries between wideband and narrowband}} systems are derived, while an efficient \textbf{\textit{wideband precoding}} scheme is also developed to enhance SE for U6G XL-MIMO systems.

\section{Wideband Boundaries from Near-Field and Far-Field Perspectives}\label{Sec: Wideband Boundary}

The Rayleigh distances in \cite{lens_far} and \cite{beamsquint_bound} have been redefined with the aim of accounting for the effect of beam squint and guiding the design of hybrid precoding scheme. However, the solutions are derived based on a directional beamforming vector, which cannot be directly used for U6G XL-MIMO systems due to the presence of multiple paths. Moreover, the near-field and far-field paths may coexist in U6G XL-MIMO channels, and thus the effect of hybrid-field paths should be given due consideration. Consequently, precise \textbf{\textit{wideband boundaries}} that do not rely on a specific beamforming vector are now being explored from near-field and far-field perspectives, respectively. The solutions can then be used to guide the specific design of efficient wideband precoding.

\subsection{Wideband Boundary in Near-Field Region}

It is important to note that the beam squint matrix of the $l$-th near-field path in {\bl\eqref{Eq: Section2_11}} has constant amplitude, i.e., $\left|\left[{\bf Q}_{\rm wn} \left( \theta_l, d_l\right) \right]_{n,m}\right| = 1$, for $n = 1,\ldots,N$, $m = 1,\ldots,M$, and $l = 1,\ldots,L$. Consequently, it can be deduced that the beam squint effect will have no influence on the power of each path. However, due to the extremely large-scale array and wide bandwidth, a phase shift will be induced for the received signal of each path at each antenna and each subcarrier.\footnote{The beam squint effect on each near/far-field path is evaluated individually, while the wideband precoding scheme that considers the presence of multiple paths will be introduced in Section $\rm \uppercase \expandafter{\romannumeral4}$.} The maximum phase shift in ${{\bf{Q}}_{{\rm{wn}}}}\left( {\theta_l, d_l} \right)$ can be denoted as
\begin{equation}\label{Eq: Section3_1_1}
\mathop {\max }\limits_{m,n} \angle {{\bf{Q}}_{{\rm{wn}}}}\left( {\theta_l, d_l} \right) = \mathop {\max }\limits_{m,n} \left| {\frac{{2\pi }}{c}{\delta _{M,m}}\Delta f\Delta d_{n,l}} \right|.
\end{equation}
Since the frequency variation and distance variation can be decoupled in each path, the formula {\bl\eqref{Eq: Section3_1_1}} can be alternatively expressed as
\begin{equation}\label{Eq: Section3_1_2}
\mathop {\max }\limits_{m,n} \angle {{\bf{Q}}_{{\rm{wn}}}}\left( {\theta_l, d_l} \right) = \frac{{2\pi }}{c}\left( {\mathop {\max }\limits_m \left| {{\delta_{M,m}} \Delta f} \right|} \right) \left( {\mathop {\max } \limits_n \left| {\Delta d_{n,l}} \right|} \right),
\end{equation}
where the maximum frequency variation is $\mathop {\max }\limits_m \left| {{\delta_{M,m}}\Delta f} \right| = \frac{B}{2}$. Subsequently, based on the spatial geometry relation, the maximum distance variation $\mathop {\max }\limits_n \left| {\Delta d_{n,l}} \right|$ is given by
\begin{equation}\label{Eq: Section3_1_3}
\mathop {\max }\limits_{n,l} \left| {\Delta d_{n,l}} \right| = \left\{ {\begin{array}{*{20}{l}}
{d_{N,l} - d_l,\,\,\, -1 < \theta < 0}\\
{d_{1,l} - d_l,\,\,\,\,\, 0 \le \theta < 1},
\end{array}} \right.
\end{equation}
and, thus, the maximum distance variation in the near-field region, $f_{\rm wn}\left(\theta, d \right) = \mathop {\max }\limits_n \left| {\Delta d_{n,l}} \right|$, is given by
\begin{equation}\label{Eq: Section3_1_4}
\begin{aligned}
f_{\rm wn}\left(\theta_l,d_l\right) = &\sqrt {{d_l^2} + 2d_l\frac{N - 1}{2}s\left|\theta_l\right| + \frac{{{{\left( {N-1} \right)}^2}}}{4}{s^2}}  - d_l \\= & \sqrt{d_l^2 + \frac{\left(N-1\right)d_l\lambda_{\rm c}\left|\theta_l\right|}{2} + \frac{\left(N-1\right)^2\lambda_{\rm c}^2}{16}} - d_l.
\end{aligned}
\end{equation}
Substituting {\bl\eqref{Eq: Section3_1_4}} into {\bl\eqref{Eq: Section3_1_2}}, the maximum phase shift in ${\bf Q}_{\rm wn}\left(\theta_l,d_l\right)$ can be calculated according to
\begin{equation}\label{Eq: Section3_1_5}
\mathop {\max }\limits_{m,n} \angle {{\bf{Q}}_{{\rm{wn}}}}\left( {\theta_l, d_l} \right) = \frac{\pi B}{c}f_{\rm wn}\left(\theta_l,d_l\right).
\end{equation}

Note that to distinguish between near-field and far-field paths, the type of a path can be determined by the maximum phase shift caused by the extremely large-scale array. In particular, if the maximum phase shift across the entire array is less than $\frac{\pi}{8}$, the spherical wave path model can be simplified to be the planar wave path model \cite{near-field/far-field}. In a similar manner, the effect of beam squint on a path is also evaluated by the maximum phase shift in ${\bf Q}_{\rm wn}\left(\theta_l, d_l \right)$, and thus the analysis is independent of a specific beamforming vector.

\textbf{\textit{Theorem 1:}} In the near-field region, the wideband boundary observed from the frequency domain can be denoted as
\begin{equation}\label{Eq: Section3_1_9}
{\bar B}_{\rm wn} = \frac{\left( \kappa_{\rm a} + \kappa_{\rm f} \right) c} {f_{\rm wn} \left(\theta_l,d_l\right)},
\end{equation}
where $\kappa_{\rm a}$ and $\kappa_{\rm f}$ are constants, while the wideband boundary observed from the antenna domain can be calculated as
\begin{equation}\label{Eq: Section3_1_10}
{\bar N}_{\rm wn} = \frac{-{\cal A}_2+\sqrt{{\cal A}_2^2+4{\cal A}_1{\cal A}_3}}{2{\cal A}_1}+1,
\end{equation}
where ${\cal A}_1 = \frac{\lambda_{\rm c}^2}{16}$, ${\cal A}_2 = \frac{\lambda_{\rm c}d_l\left|\theta_l\right|}{2}$, and ${\cal A}_3 = \frac{\left( \kappa_{\rm a} + \kappa_{\rm f} \right)^2 c^2+2\left( \kappa_{\rm a} + \kappa_{\rm f} \right)cd_lB}{B^2}$.

\textbf{\textit{Proof:}} See Appendix A.

For a U6G XL-MIMO system, when $N$ is constant, if the bandwidth satisfies $B < {\bar B}_{\rm wn}$, or when $B$ is constant, if the number of antennas satisfies $N < {\bar N}_{\rm wn}$, the beam squint effect on this path can be ignored when we design the hybrid precoder in the near-field perspective, such that
\begin{equation}\label{Eq: Section3_1_7}
{\bf Q}_{\rm wn} \left( {\theta_l, d_l} \right) \approx {{\bf{1}}_N} \otimes {{\bf{1}}_M^{\rm T}}.
\end{equation}
Hence, the wideband near-field (WN) path model as in {\bl\eqref{Eq: Section2_10}} can be simplified to be a narrowband near-field (NN) path model in \cite{narrowband_model}, ${\bf H}_{{\rm nn},l} \in \mathbb{C}^{N \times M}$, which is given by
\begin{equation}\label{Eq: Section3_1_8}
{\bf H}_{{\rm nn},l} = {g_l {\bf{w}}\left( {{\theta_l },{d_l}} \right)} {{\bf{p}}^{\rm{T}}}\left( {d_l, r_l} \right).
\end{equation}
Then, conventional near-field transmission schemes remain applicable due to the negligible effect of beam squint.

\begin{figure}
  \centering
  \includegraphics[scale=0.29]{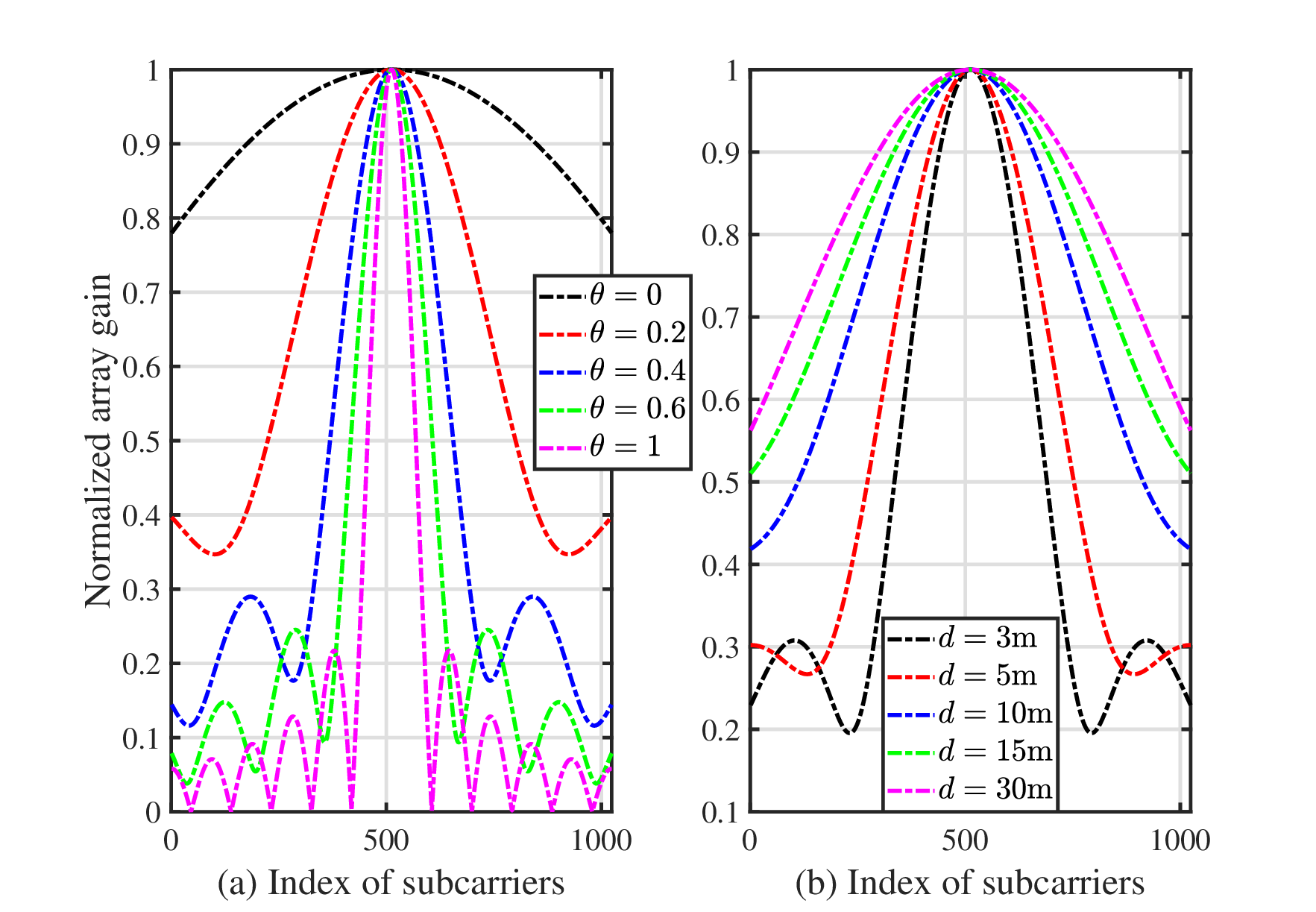}
  \caption{Normalized array gain $\eta_{\rm wn}\left(m,\theta,d\right)$ as \eqref{Eq: Section3_3_3} against $\theta$ and $d$ ($f_{\rm c} = 7$ GHz, $B = 300$ MHz, $M = 1024$, $N = 512$, $\theta = 0.1$, and $d = r = 20$ m).} \label{Fig: beam squint effect}
\end{figure}

\textbf{\textit{Proposition 1:}} From a near-field perspective, it is evident that the antenna-domain and frequency-domain wideband boundaries, i.e. ${\bar B}_{\rm wn}$ and ${\bar N}_{\rm wn}$, both decrease with an increase in $\left|\theta_l\right|$. As shown in Fig.~\ref{Fig: beam squint effect}(a), the beam squint effect is minimal in the normal direction, while it becomes pronounced as the incident signal deviates from the array normal.

\textbf{\textit{Proof:}} For the near-field frequency-domain wideband boundary $\bar B_{\rm wn}$, it is obvious that $f_{\rm wn}\left(\theta_l,d_l\right)$ is a monotonically increasing function w.r.t. $\left|\theta_l\right|$ as in {\bl\eqref{Eq: Section3_1_4}}, and, thus, $\bar B_{\rm wn}$ is a monotonically decreasing function w.r.t. $\left|\theta_l\right|$ as in {\bl\eqref{Eq: Section3_1_9}}.

Since the near-field antenna-domain wideband boundary $\bar N_{\rm wn}$ has been denoted as {\bl\eqref{Eq: Section3_1_10}}, it can be alternatively written as
\begin{equation}\label{Eq: Section3_1_13}
\bar N_{\rm wn} = \frac{- {\cal A}_4\left|\theta_l\right| + \sqrt{{\cal A}_4^2\theta_l^2 + 4{\cal A}_1{\cal A}_3}}{2{\cal A}_1} + 1 = \frac{\chi\left(\theta_l\right)}{2{\cal A}_1}+1,
\end{equation}
where ${\cal A}_3 > 0$, and ${\cal A}_4 = \frac{\lambda_{\rm c}d_l}{2} > 0$. Note that the first-order derivative of $\chi\left(\theta_l\right)$ for $0 \le \theta_l < 1$ satisfies
\begin{equation}\label{Eq: Section3_1_14}
\frac{\partial \chi\left(\theta_l\right)}{\partial \theta_l} = \frac{{\cal A}_4^2}{\sqrt{\frac{4{\cal A}_1{\cal A}_3}{\theta_l^2}+{\cal A}_4^2}} - {\cal A}_4,
\end{equation}
which is a monotonically increasing function w.r.t. $\theta_l$, and satisfies $\frac{\partial \chi\left(\theta_l\right)}{\partial \theta_l}|_{\theta_l = 1} = \frac{{\cal A}_4^2}{\sqrt{4{\cal A}_1{\cal A}_3+{\cal A}_4^2}} - {\cal A}_4 < 0$. Thus, $\frac{\partial \chi\left(\theta_l\right)}{\partial \theta_l} < 0$ always holds for $0 \le \theta_l < 1$, and $\chi \left(\theta_l\right)$ is a monotonically decreasing function w.r.t. $\theta_l$. Then, ${\bar N}_{\rm wn}$ decreases with an increase in $\left|\theta_l\right|$. Hence, \textbf{\textit{Proposition 1}} is proved. \qed

\textbf{\textit{Proposition 2:}} From a near-field perspective, the frequency-domain wideband boundary $\bar B_{\rm wn}$ has a lower bound $\bar B_{\rm wn}^{\rm lower} = \frac{4\left( \kappa_{\rm a} + \kappa_{\rm f} \right)f_{\rm{c}}}{N-1}$, which is achieved when $\left|\theta_l \right| \to 1$, while an upper bound $\bar B_{\rm wn}^{\rm upper} = \frac{4 \left(\kappa_{\rm a} + \kappa_{\rm f} \right) c}{\sqrt{\lambda_{\rm c}^2 \left(N - 1\right)^2 + 16d_l^2} - 4d_l}$ is achieved when $\theta_l = 0$, respectively. Similarly, the antenna-domain wideband boundary also has a lower bound $\bar N_{\rm wn}^{\rm lower} = \frac{4 \left(\kappa_{\rm a} + \kappa_{\rm f} \right) f_{\rm c}}{B} + 1$ ($\left|\theta_l\right| \to 1$), and an upper bound $\bar N_{\rm wn}^{\rm upper} = \frac{\sqrt{{\cal A}_1 {\cal A}_3}}{{\cal A}_1} + 1$ ($\theta_l = 0$), which determine the maximum acceptable bandwidth and array size to make the beam squint effect negligible.

\textbf{\textit{Proof:}} Based on \textbf{\textit{Proposition 1}}, the upper and lower bounds of the maximum distance variation can be computed as
\begin{equation}\label{Eq: Section3_1_15}
\left\{ {\begin{array}{*{20}{ll}}
{\mathop {\max }\limits_{\theta_l}  f_{\rm wn}\left(\theta_l,d_l\right) = \frac{\left(N - 1\right){\lambda _{\rm{c}}}}{4}, \,\,\,\, \,\,\,\, \,\,\,\, \,\,\,\, \,\,\,\, \,\,\,\, \,\,\,\, \,\,\, \,\, \left|\theta_l\right| \to 1}\\
{\mathop {\min }\limits_{\theta_l} f_{\rm wn}\left(\theta_l,d_l\right) \mathop = \frac{\sqrt{\lambda_{\rm c}^2 \left(N - 1\right)^2 + 16d_l^2}}{4} - d_l,\,\,\,\theta_l = 0}.
\end{array}} \right.
\end{equation}
Substituting {\bl\eqref{Eq: Section3_1_15}} into {\bl\eqref{Eq: Section3_1_9}} and {\bl\eqref{Eq: Section3_1_10}}, and then \textbf{\textit{Proposition 2}} is proved. \qed

\textbf{\textit{Discussion:}} A U6G XL-MIMO system is evaluated, where $f_{\rm c} = 7$ GHz, $N = 512$, $d = 40$ m, and $\kappa_{\rm a} = \kappa_{\rm f} = 0.125$. Then, $B_{\rm wn}^{\rm lower} \approx 13.7$ MHz, and $\bar B_{\rm wn}^{\rm upper} \approx 200$ MHz for a near-field path, which are typically smaller than $B$ in the U6G frequency band (where the maximum bandwidth is up to $700$ MHz). When $B = 300$ MHz, and $\theta = 0.3$, such that $\bar N_{\rm wn} \approx 58$, which typically satisfies $\bar N_{\rm wn} \ll N$ in XL-MIMO systems. Therefore, the issue of beam squint effect is still pronounced in the U6G XL-MIMO systems.

\subsection{Wideband Boundary in Far-Field Region}

As demonstrated in \cite{near-field/far-field}, when the maximum phase shift variation of the $l$-th path across the entire array is smaller than a threshold $\kappa_{\rm a} \pi$, the near-field spherical wave steering vector of this path can be simplified to be a far-field planar wave steering vector, for $l = 1,\ldots,L$. The antenna-domain steering vector of the $l$-th far-field path, ${\bf a}\left(\theta_l \right) \in \mathbb{C}^{N \times 1}$, can be given by
\begin{equation}\label{Eq: Section3_1_16}
{\bf{a}}\left(\theta_l \right) = {\left[ {{e^{j\frac{{2\pi }}{c}{f_{\rm{c}}}{\delta _{N,1}}s\theta_l }}, \cdots ,{e^{j\frac{{2\pi }}{c}{f_{\rm{c}}}{\delta _{N,N}}s\theta_l }}} \right]^{\rm{T}}},
\end{equation}
while ${{\bf{Q}}_{{\rm{wn}}}}\left( {\theta_l ,d_l} \right)$ can be simplified to be a far-field beam squint matrix ${{\bf{Q}}_{{\rm{wf}}}}\left( {\theta_l} \right) \in {\mathbb{C}^{N \times M}}$, which can be expressed as
\begin{equation}\label{Eq: Section3_1_17}
\left[ {{\bf Q}_{\rm wf} \left( {\theta_l} \right)} \right]_{n,m} = e^{j\frac{{2\pi }}{c}{\delta _{M,m}}\Delta f{\delta _{N,n}}s\theta_l},
\end{equation}
for $n = 1,\ldots,N$, and $m = 1,\ldots,M$. Consequently, when the beam squint effect is pronounced, the wideband far-field (WF) path model ${\bf H}_{{\rm wf},l} \in \mathbb{C}^{N \times M}$ in \cite{beamsquint1} can be denoted as
\begin{equation}\label{Eq: Section3_1_18}
{\bf H}_{{\rm wf},l} = g_l \left({\bf a}\left( \theta_l \right){{\bf{p}}^{\rm T}}\left( {d_l,r_l} \right)\right) \odot {\bf Q}_{\rm wf}\left( {\theta_l} \right).
\end{equation}

\textbf{\textit{Theorem 2:}} From the far-field perspective, the wideband boundary observed from the frequency domain can be determined as
\begin{equation}\label{Eq: Section3_1_19}
{\bar B}_{\rm wf} = \frac{\left( \kappa_{\rm a} + \kappa_{\rm f} \right) c}{f_{\rm wf} \left(\theta_l \right)},
\end{equation}
while the wideband boundary observed from the antenna domain can be calculated as
\begin{equation}\label{Eq: Section3_1_20}
{\bar N}_{\rm wf} = \frac{4 \left( \kappa_{\rm a} + \kappa_{\rm f} \right) {f_{\rm{c}}}} {B\left| \theta_l \right|} + 1.
\end{equation}

\textbf{\textit{Proof:}} See Appendix B.

In an XL-MIMO system, if the bandwidth satisfies $B < {\bar B}_{\rm wf}$ when $N$ is constant, or the number of antennas satisfies $N < \bar N_{\rm wf}$ when $B$ is constant, the beam squint from the far-field perspective can be ignored, i.e. ${\bf Q}_{\rm wf} \left(\theta_l\right) \approx {\bf 1}_N \otimes {\bf 1}_M^{\rm T}$, while the conventional far-field transmission scheme performs perfectly. The WF path model can be simplified to be a narrowband far-field (NF) path model \cite{narrowband_model}, ${\bf{H}}_{{\rm nf},l} \in \mathbb{C}^{N \times M}$, such that
\begin{equation}\label{Eq: Section3_1_22}
{\bf{H}}_{{\rm nf},l} = g_l{\bf{a}}\left( {\theta_l} \right) {{\bf p}^{\rm{T}}} \left( {d_l,r_l} \right).
\end{equation}

\textbf{\textit{Proposition 3:}} Similarly to the solution in the near-field perspective, the effect of the beam squint in the far-field region is also amplified with an increase in $\left|\theta_l\right|$, since ${\bar B}_{\rm wf}$ and ${\bar N}_{\rm wf}$ are both monotonically decreasing functions w.r.t. $\left| \theta_l \right|$ as in {\bl\eqref{Eq: Section3_1_20}} and {\bl\eqref{Eq: Section3_1_21}}. This solution means that the beam squint in the far-field region also becomes pronounced when the incident signal deviates from the normal direction of the array.

\begin{table*}[htb]   
\begin{center}   
\caption{Three Types of Path Models and Boundaries between Wideband/Narrowband Near/Far-Field Channels}
\label{Tab: channel types} 
\begin{tabular}{c|c|c|c|c}   
\hline
\textbf{Models} & \textbf{Threshold} & \textbf{Boundaries} & \textbf{Lower bounds} & \textbf{Upper bounds} \\
\hline
\multirow{2}*{WN {\bl \eqref{Eq: Section2_10}}} & $N \ge {\bar N}_{\rm wn}$ & \multirow{2}*{$\bar N_{\rm wn} = \frac{-{\cal A}_2 + \sqrt{{\cal A}_2^2 + 4{\cal A}_1{\cal A}_3}}{2{\cal A}_1} + 1$} & \multirow{2}*{$\bar N_{\rm wn}^{\rm lower} = \frac{4 \left( \kappa_{\rm a} + \kappa_{\rm f} \right) f_{\rm{c}}}{B} + 1$} & \multirow{2}*{$\bar N_{\rm wn}^{\rm upper} = \frac{\sqrt{{\cal A}_1 {\cal A}_3}}{{\cal A}_1} + 1$} \\

\cline{2-2} & $B \ge \bar B_{\rm wn}$ &  &  \\

\cline{1-2} \multirow{2}*{NN {\bl \eqref{Eq: Section3_1_8}}} & $\tilde N \le N < {\bar N}_{\rm wn}$ & \multirow{2}*{$\bar B_{\rm wn} = \frac{\left(\kappa_{\rm a} + \kappa_{\rm f} \right)c}{f_{\rm wn} \left(\theta,d\right)}$} & \multirow{2}*{$\bar B_{\rm wn}^{\rm lower} = \frac{4 \left( \kappa_{\rm a} + \kappa_{\rm f} \right) f_{\rm{c}}}{N - 1}$} & \multirow{2}*{$\bar B_{\rm wn}^{\rm upper} = \frac{4 \left( \kappa_{\rm a} + \kappa_{\rm f} \right) c}{\sqrt{\lambda_{\rm c}^2 {\left( {N - 1} \right)}^2 + 16 d^2} - d}$} \\

\cline{2-2} & $B < \bar B_{\rm wn}$ &  &  \\

\cline{1-2} {NF {\bl \eqref{Eq: Section3_1_22}}} & $N < \tilde N$ & ${\tilde N} = \frac{-{\cal A}_2 + \sqrt{{\cal A}_2^2 + 4{\cal A}_1 {\cal A}_5}}{2{\cal A}_1} + 1$ & {$\tilde N^{\rm lower} = 2\kappa_{\rm a} + 1$} & {$\tilde N^{\rm upper} = \frac{\sqrt{{\cal A}_1 {\cal A}_5}}{{\cal A}_1} + 1$} \\
\hline
\end{tabular}
\end{center}
\end{table*}

\textbf{\textit{Proposition 4:}} From the far-field perspective, the frequency-domain wideband boundary has a lower bound $\bar B_{\rm wf}^{\rm lower} = \frac{4 \left(\kappa_{\rm a} + \kappa_{\rm f} \right) f_{\rm{c}}}{N-1}$, when $\left|\theta_l\right| \to 1$, and an upper bound $\bar B_{\rm wf}^{\rm upper} \to \infty$, when $\theta_l = 0$. The antenna-domain wideband boundary also has a lower bound $\bar N_{\rm wf}^{\rm lower} = \frac{4 \left(\kappa_{\rm a} + \kappa_{\rm f} \right) f_{\rm{c}}}{B} + 1$ when $\left|\theta_l\right| \to 1$, and an upper bound $\bar N_{\rm wf}^{\rm upper} \to \infty$ when $\theta_l = 0$, respectively. This implies that whenever the incident signal comes from the array normal direction in the far-field region, the beam squint effect can be ignored, even $B$ and $N$ tend to infinity.

\textbf{\textit{Proposition 5:}} The beam squint is more pronounced in near-field regions than in far-field regions, i.e., $\bar B_{\rm wn} < \bar B_{\rm wf}$ and $\bar N_{\rm wn} < \bar N_{\rm wf}$, for $-1 < \theta_l < 1$, while $\bar B_{\rm wn} = \bar B_{\rm wf}$ and $\bar N_{\rm wn} = \bar N_{\rm wf}$, only when $\left|\theta_l\right| \to 1$. Thus, it can be deduced that beam squint is imperative to be considered as an issue in U6G XL-MIMO systems, due to the inherent near-field effect.

\textbf{\textit{Proof:}} Since the frequency-domain wideband boundaries in the near-field and far-field regions can be expressed as {\bl\eqref{Eq: Section3_1_9}} and {\bl\eqref{Eq: Section3_1_19}}, respectively, the maximum distance variation function satisfies $f_{\rm wn}\left(\theta_l,d_l\right) \mathop \approx \limits^{\left( {\rm{a}} \right)} f_{\rm wf}\left(\theta_l\right) + \frac{{{{\left( {N - 1} \right)}^2}\lambda_{\rm{c}}^2 \left( {1 - {\theta_l ^2}} \right)}}{{32d_l}}$, where $\left({\rm a}\right)$ is derived via $\sqrt{1+x} \approx 1+\frac{1}{2}x-\frac{1}{8}x^2$. Note that $f_{\rm wn}\left(\theta_l,d_l\right) > f_{\rm wf} \left(\theta_l\right)$ holds for $-1< \theta_l < 1$, and $f_{\rm wn}\left(\theta_l,d_l\right) = f_{\rm wf}\left(\theta_l\right)$, only when $\left|\theta_l\right| \to 1$. Then, we can get $\bar B_{\rm wn} < \bar B_{\rm wf}$ for $-1 < \theta_l < 1$, and $\bar B_{\rm wn} = \bar B_{\rm wf}$, only when $\left|\theta_l\right| \to 1$.

Note that the antenna-domain wideband boundary satisfies $f_{\rm wn}\left(\theta_l,d_l\right) \approx f_{\rm wf}\left(\theta_l\right)$ when $\left|\theta_l\right| \to 1$, and then $\bar N_{\rm wn} = \bar N_{\rm wf}$ ($\left|\theta_l\right| \to 1$). For $-1 < \theta_l < 1$, let $\mathop {\max }\limits_{m,n} \angle {\bf Q}_{\rm wn}\left(\theta_l,d_l\right) = \mathop {\max }\limits_{m,n} \angle {\bf Q}_{\rm wf}\left(\theta_l\right)$, which can be alternatively expressed as
\begin{equation}\label{Eq: Section3_1_23}
f_{\rm wn}\left(\theta_l,d_l\right)|_{N = \bar N_{\rm wn}} = f_{\rm wf}\left(\theta_l\right)|_{N = \bar N_{\rm wf}}.
\end{equation}
Since $f_{\rm wn}\left(\theta_l,d_l\right)$ and $f_{\rm wf}\left(\theta_l\right)$ are both monotonically increasing functions w.r.t. $N$, $f_{\rm wn}\left(\theta_l,d_l\right) > f_{\rm wf}\left(\theta_l\right)$ always holds for the same $N$. Thus, we can get $\bar N_{\rm wn} < \bar N_{\rm wf}$ through {\bl\eqref{Eq: Section3_1_23}}. \qed

\textbf{\textit{Proposition 6:}} As shown in Fig.~\ref{Fig: beam squint effect}(b), the beam squint has a stronger effect with the UE moving closer to the BS, i.e., $\bar B_{\rm wn}$ and $\bar N_{\rm wn}$ both reduce when $d_l$ decreases.

\textbf{\textit{Proof:}} Since $f_{\rm wn}\left(\theta_l,d_l\right) \approx f_{\rm wf}\left(\theta_l\right) + \frac{\left(N-1\right)^2\lambda_{\rm c}^2\left(1-\theta_l^2\right)}{32d_l}$, in which $\frac{\left(N-1\right)^2\lambda_{\rm c}^2\left(1-\theta_l^2\right)}{32d_l}$ is a monotonically decreasing function w.r.t. $d_l$, where $f_{\rm wf}\left(\theta_l\right)$ is considered constant. Thus, $\bar B_{\rm wn}$ as in {\bl\eqref{Eq: Section3_1_9}} is a monotonically increasing function of $d_l$.

In the antenna domain, the first-order derivative w.r.t. $d_l$ for $f_{\rm wn}\left(\theta_l,d_l\right)$ in {\bl\eqref{Eq: Section3_1_11}}, satisfies
\begin{equation}\label{Eq: Section3_1_24}
\frac{\partial f_{\rm wn}\left(\theta_l,d_l\right)} {\partial d_l} = \frac{1}{\sqrt{\frac{{\cal A}_1\left(N-1\right)^2 + {\cal A}_2\left(N-1\right)}{d_l^2} + 1}} - 1 < 0,
\end{equation}
and, thus $f_{\rm wn}\left(\theta_l,d_l\right)$ is a monotonically decreasing function of $d_l$, where $\theta_l$ is a constant, while $f_{\rm wn} \left(\theta_l,d_l\right)$ is also a monotonically increasing function of $N$. Hence, substituting {\bl\eqref{Eq: Section3_1_5}} into {\bl\eqref{Eq: Section3_1_6}}, then $f_{\rm wn} \left(\theta_l,d_l\right) < \frac{\left(\kappa_{\rm a} + \kappa_{\rm f}\right) c}{B}$. When $d_l$ decreases, a larger $\bar N_{\rm wn}$ can be obtained. This concludes the proof. \qed

Note that the frequency-domain and antenna-domain wideband boundaries have the same lower bounds in the near-field and far-field regions, i.e., $\bar B_{\rm wn}^{\rm lower} = \bar B_{\rm wf}^{\rm lower} = \frac{4\left( \kappa_{\rm a} + \kappa_{\rm f} \right) f_{\rm c}}{N-1}$, and $\bar N_{\rm wn}^{\rm lower} = \bar N_{\rm wf}^{\rm lower} = \frac{4 \left( \kappa_{\rm a} + \kappa_{\rm f} \right) f_{\rm c}}{B} + 1$. In practical systems, it is challenging to obtain perfect channel state information in a timely manner, due to the effects of multiple moving UEs, multipath, and hybrid-field features. In this instance, the lower bounds can still be used as a general indicator to determine whether the beam squint is pronounced, since they are independent of specific channel parameters.

\textbf{\textit{Discussion:}} Compared to mmWave and THz systems, a smaller array and bandwidth are constrained for U6G XL-MIMO systems to mitigate the beam squint effect, since both $\left\{\bar B_{\rm wn},\bar B_{\rm wf}\right\}$ are proportional to $f_{\rm c}$ when $N$ is a constant. In addition, both $\left\{\bar N_{\rm wn},\bar N_{\rm wf}\right\}$ are also proportional to $f_{\rm c}$ when $B$ is a constant.

\textbf{\textit{Theorem 3:}} When $B$ is constant, the manifestation of the near-field effect precedes that of the beam squint effect as $N$ increases, such that
\begin{equation}\label{Eq: Section3_1_25}
{\tilde N} < \min \left\{ \bar N_{\rm wn}, \bar N_{\rm wf} \right\}.
\end{equation}
Moreover, when $N$ is constant, the manifestation of the near-field effect still precedes that of the beam squint effect as $B$ increases. The following formula always holds
\begin{equation}\label{Eq: Section3_1_26}
B < \min \left\{ \bar B_{\rm wn}, \bar B_{\rm wf} \right\},
\end{equation}
since $B < f_{\rm c}$. It can be concluded that the beam squint effect is invariably negligible in the far-field region, while it becomes pronounced only in the near-field region.

\textbf{\textit{Proof:}} See Appendix C.

Hence, the paths can be summarized into three types in practical systems, i.e., the WN, NN, and NF models. The three types of path models and their boundaries have been summarized in \textbf{Table~\ref{Tab: channel types}}, where the derivation processes of $\tilde N^{\rm lower}$ and $\tilde N^{\rm upper}$ are similar to \textbf{\textit{Proposition 1}} and \textbf{\textit{Proposition 2}}.

\subsection{Wideband Boundary Analysis}

\begin{figure}
  \centering
  \includegraphics[scale=0.29]{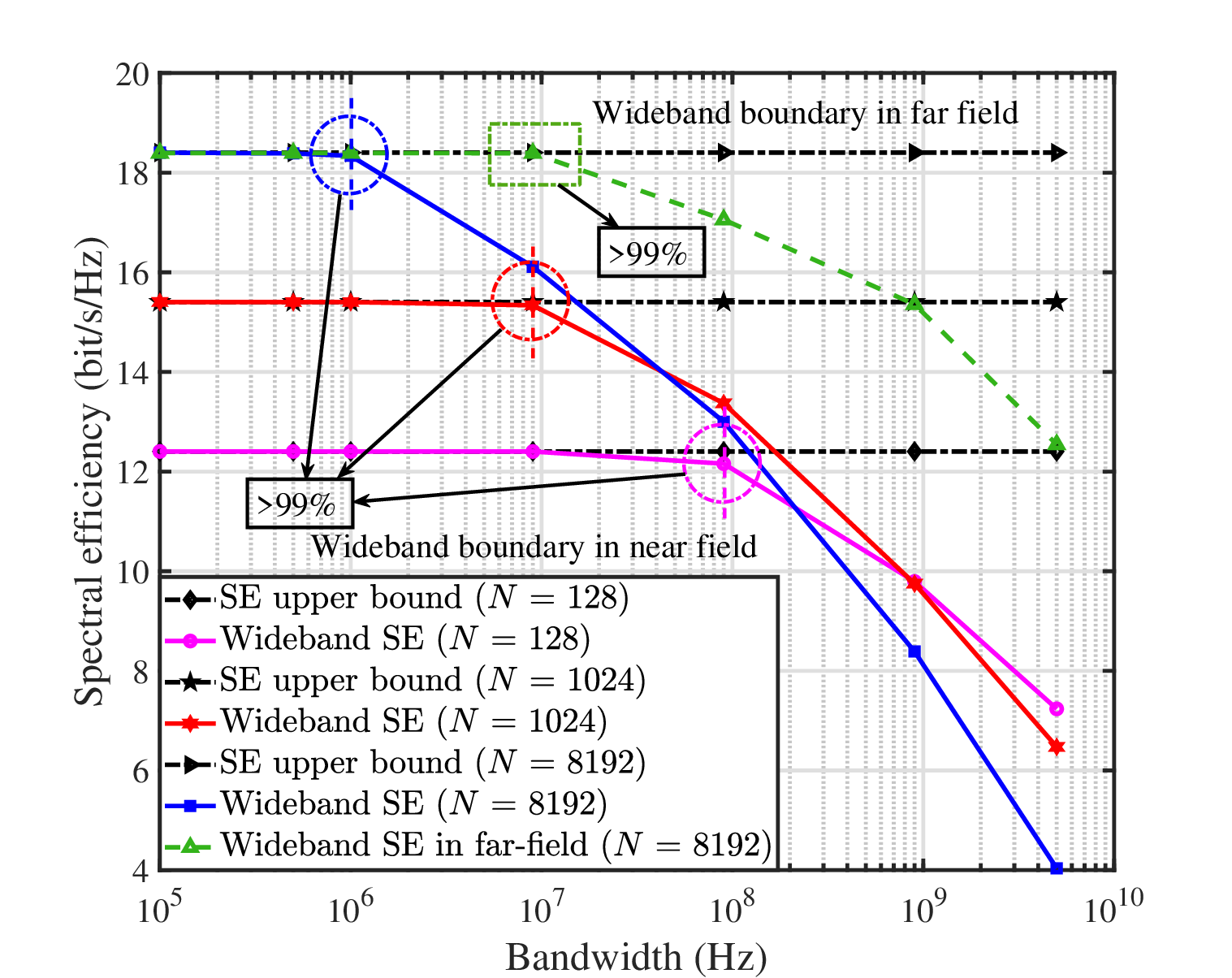}
  \caption{SE against the bandwidth for U6G XL-MIMO systems.} \label{Fig: SE with B}
  \vspace{-0.4cm}
\end{figure}

To evaluate the derived wideband boundaries, a single-path scenario is considered firstly, while the design of wideband precoding for hybrid-field multipath scenarios is introduced in Section $\rm \uppercase \expandafter{\romannumeral4}$. A frequency-independent maximum ratio transmission (MRT) precoding is designed as ${\bf f}_{\rm wn} \in \mathbb{C}^{N \times 1}$, which points to the position of the UE/scatterer, such that
\begin{equation}\label{Eq: Section3_3_1}
{\bf f}_{\rm wn} = \frac{1}{\sqrt{N}}{\bf w}\left(\theta,d\right).
\end{equation}
The variance of noise is indicated by $\sigma^2$, and then the SE is
\begin{equation}\label{Eq: Section3_3_2}
\begin{aligned}
&{\rm{SE}}_{{\rm{wn}}} = \frac{1}{M} \sum\limits_{m = 1}^M {\log _2}\left( {1 + \frac{P{{\left| {{\bf f}_{\rm wn}^{\rm H}\left[{\bf H}_{\rm wn}\right]_{:,m}} \right|}^2}}{\sigma ^2}} \right) \\= &\frac{1}{M} \sum\limits_{m = 1}^M {\log_2}\left( {1 + \frac{PN{{\left| g \right|}^2} {\eta_{\rm wn}^2 \left({m,\theta ,d}\right)}}{\sigma ^2}} \right),
\end{aligned}
\end{equation}
where $P$ is the transmit power, while ${\eta}_{\rm wn} \left( {m,\theta ,d} \right)$ denotes the normalized array gain, which is given by
\begin{equation}\label{Eq: Section3_3_3}
\eta_{\rm wn} \left( {m,\theta ,d} \right) = \frac{\left| {\bf f}_{\rm wn}^{\rm H} \left[ {\bf H}_{\rm wn} \right]_{:,m} \right|}{\sqrt{N}} = \frac{{\left| {\sum\limits_{n = 1}^N {{{\left[ {{{\bf{q}}_{\rm{c}}}\left( {m,\theta ,d} \right)} \right]}_n}} } \right|}}{N},
\end{equation}
for $m = 1,\ldots,M$. Note that when the beam squint is non-negligible in U6G XL-MIMO systems, and since
\begin{equation}\label{Eq: Section3_3_4}
{\left| {\sum\limits_{n = 1}^N {{{\left[ {{{\bf{q}}_{\rm{c}}}\left( {m,\theta ,d} \right)} \right]}_m}} } \right|} \le \sum\limits_{n = 1}^N {\left| {{{\left[ {{{\bf{q}}_{\rm{c}}}\left( {m,\theta ,d} \right)} \right]}_m}} \right|}  = N,
\end{equation}
the normalized array gain satisfies $\mathop {\max }\limits_{m,\theta ,d} {\eta_{\rm wn} \left( {m,\theta ,d} \right)} \le 1$ (where $\mathop {\max }\limits_{m,\theta ,d} {\eta_{\rm wn} \left( {m,\theta ,d} \right)} = 1$ only when ${\delta _{M,m}}\Delta f = 0$). Hence, the SE of the wideband near-field channel satisfies
\begin{equation}\label{Eq: Section3_3_5}
{\rm{SE}}_{{\rm{wn}}} < \frac{1}{M} \sum\limits_{m = 1}^M {\log _2}\left( {1 + \frac{PMN{{\left| g \right|}^2}}{\sigma ^2}} \right) = {\rm SE}_{\rm opt},
\end{equation}
where ${\rm SE}_{\rm opt} = {\log _2}\left( {1 + \frac{{PMN{{\left| g \right|}^2}}}{{{\sigma ^2}}}} \right)$ is the SE upper bound with a fully digital precoding architecture, which can be written as
\begin{equation}\label{Eq: Section3_3_6}
{\rm{SE}}_{{\rm{opt}}} = \frac{1}{M}\sum\limits_{m = 1}^M{\log _2}\left( {1 + \frac{P {{\left| {{\bf f}_{{\rm opt},m}^{\rm H}{{\left[ {{\bf H}_{\rm wn}} \right]}_{:,m}}} \right|}^2}}{\sigma ^2}} \right),
\end{equation}
where ${{\bf{f}}_{{\rm{opt}},m}} \in {\mathbb{C}^{N \times 1}}$ is the frequency-dependent optimal receiving vector, where the beams at $M$ subcarriers all point to the UE, such that
\begin{equation}\label{Eq: Section3_3_7}
{{\bf{f}}_{{\rm{opt}},m}} = \frac{1}{\sqrt{N}} {{\bf{w}}\left( {\theta ,d} \right) \odot {{\bf{q}}_{\rm{c}}}\left( {m,\theta ,d} \right)}.
\end{equation}

The SE against $B$ is simulated for U6G XL-MIMO systems, where $f_{\rm c} = 7$ GHz; $M = 1024$; $\theta = 0.1$; $d = r = 10$ m, and $\kappa_{\rm a} = \kappa_{\rm f} = 0.125$. The signal-to-noise (SNR) is set as 10 dB, where the SNR is defined as
\begin{equation}\label{Eq: Section3_3_8}
{\rm{SNR}} = 10{\log _{10}}\left( {\frac{{P{{\left| g \right|}^2}}}{{{\sigma ^2}}}} \right).
\end{equation}

\begin{figure}
  \centering
  \includegraphics[scale=0.29]{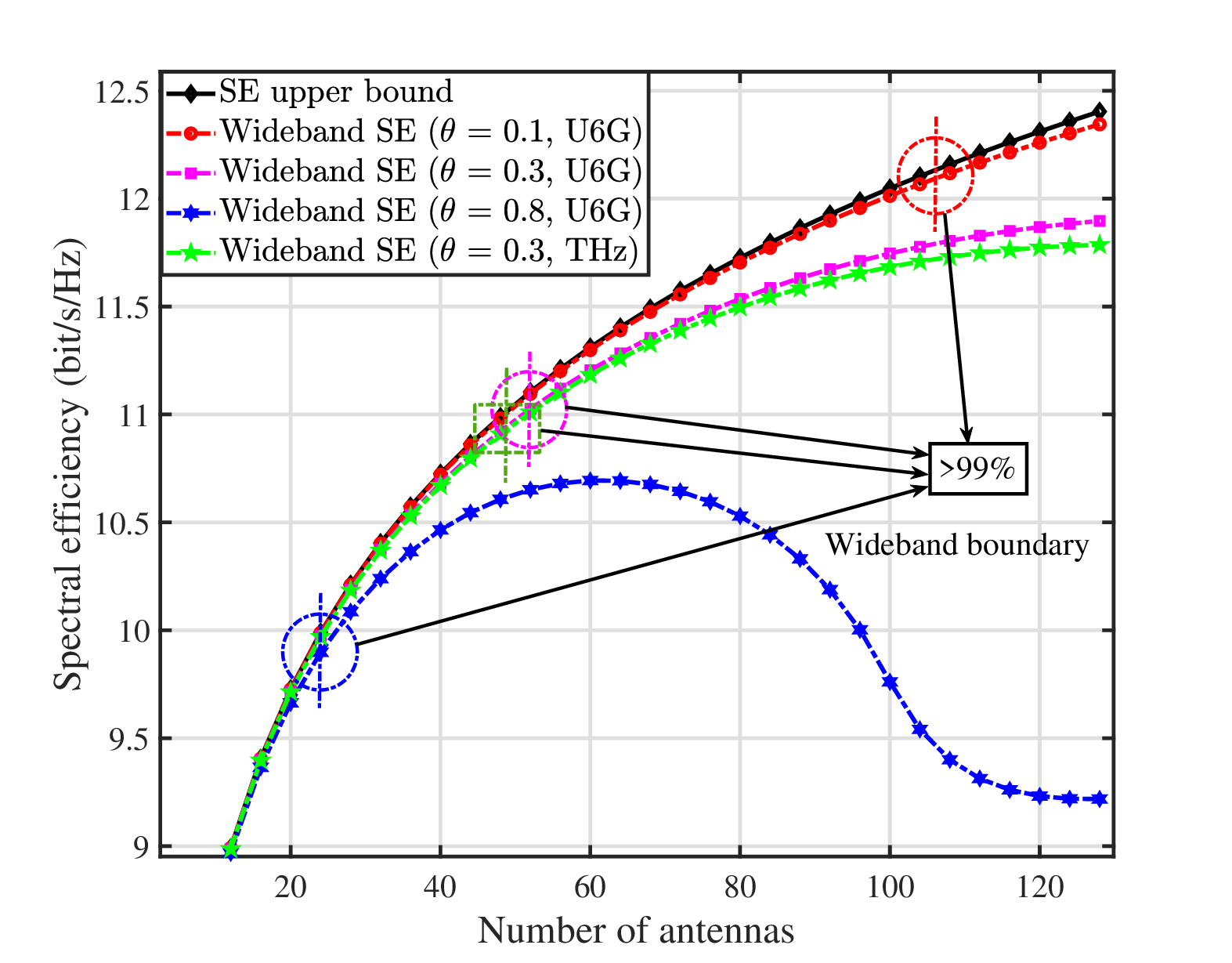}
  \caption{SE against the number of antennas for U6G XL-MIMO systems.} \label{Fig: SE with N}
\end{figure}

As shown in Fig.~\ref{Fig: SE with B}, the proposed wideband boundaries are capable of accurately characterizing wideband and narrowband conditions. Three XL-MIMO settings are considered, where the number of antennas is $N = 128$, $N = 1024$, and $N = 8192$, respectively. In the case of a smaller bandwidth than $\bar B_{\rm wn}$, the SE of the hybrid architecture is greater than 99$\%$ of ${\rm SE}_{\rm opt}$, indicating that the beam squint can be ignored. However, when $B \ge \bar B_{\rm wn}$, the SE decreases rapidly against $B$, and will be much below ${\rm SE}_{\rm opt}$, which means that the beam squint becomes pronounced. Note that the SE decreases more rapidly as $N$ increases, since a more pronounced beam squint will be caused by larger arrays. The wideband boundary in far-field perspective is also evaluated. We can find that when $B < \bar B_{\rm wf}$, the SE in the far-field region can achieve more than 99$\%$ of ${\rm SE}_{\rm opt}$, while the SE decreases rapidly when $B \ge \bar B_{\rm wf}$, due to the pronounced beam squint. However, in far-field scenarios, the beam squint effect can generally be ignored, given the bandwidth constraint in practical U6G systems.

Subsequently, the SE against $N$ is evaluated, where $B = 300$ MHz. As shown in Fig.~\ref{Fig: SE with N}, when $N < \bar N_{\rm wn}$, an SE that is approximately equal to ${\rm SE}_{\rm opt}$ can be obtained. However, when $N \ge \bar N_{\rm wn}$, the decrease in SE compared to ${\rm SE}_{\rm opt}$ becomes more significant with an increase in $N$. As the value of $\theta$ increases, the wideband boundaries decrease due to a more pronounced beam squint when the incident signal deviates from the array normal. Note that the proposed boundaries are also valid in other frequency bands (such as THz), where $f_{\rm c} = 300$ GHz, $B = 15$ GHz, and $N = 8192$.

\section{Channel Slicing and Wideband Precoding \\ for U6G XL-MIMO Systems}\label{Sec: Channel Slicing}

Due to the issue of beam squint in U6G XL-MIMO systems, constantly increasing the number of antennas and bandwidth is not a viable solution and will not bring continuous improvement of the SE. It is therefore essential to understand how many antennas and how much bandwidth are appropriate to improve the SE. Based on the wideband boundaries, a \textbf{\textit{channel slicing}} scheme is now proposed. In U6G XL-MIMO systems with adjustable and constant subarray architectures, the \textbf{\textit{wideband precoding}} methods will be introduced from the antenna-domain and frequency-domain perspectives, respectively.

\subsection{Antenna-Domain Subarray Slicing}

Some studies have indicated that dividing the entire array into subarrays has the potential to mitigate the beam squint effect \cite{TTD_far1,TTD_far2,TTD_far3,multiple_beam,subarray}. However, these studies require the deployment of TTD lines, resulting in an increase in hardware cost. Moreover, most studies focus only on a single-path scenario, while the inherent hybrid-field and multiple-path characteristics must be considered in U6G XL-MIMO systems. Therefore, based on the wideband boundaries, an efficient antenna-domain slicing scheme is now introduced to improve the SE for the U6G XL-MIMO systems. Assume that the channel between the UE and the BS contains $L_{\rm N}$ near-field and $L_{\rm F}$ far-field paths. According to \textbf{\textit{Theorem 3}}, the hybrid-field channel, ${\bf H} \in \mathbb{C}^{N \times M}$, can be expressed as
\begin{equation}\label{Eq: Section4_1_1}
\begin{aligned}
{\bf H} = &\sum\limits_{l_{\rm F} = 1}^{L_{\rm F}} g_{l_{\rm F}} {\bf a} \left( \theta_{l_{\rm F}} \right) {\bf p}^{\rm T} \left( d_{l_{\rm F}}, r_{l_{\rm F}} \right) \\+ &\sum\limits_{{l_{\rm N}} = 1}^{L_{\rm N}} g_{l_{\rm N}} \left( {\bf w} \left( \theta_{l_{\rm N}}, d_{l_{\rm N}} \right) {\bf p}^{\rm T} \left( d_{l_{\rm N}}, r_{l_{\rm N}} \right) \right) \odot {\bf Q}_{\rm wn} \left( \theta_{l_{\rm N}}, d_{l_{\rm N}} \right).
\end{aligned}
\end{equation}
Note that according to \textbf{\textit{Theorem 3}}, the beam squint effect in each far-field path can be ignored, while the beam squint effect in the near-field components still requires consideration to enhance the SE for U6G XL-MIMO systems. However, according to \textbf{\textit{Theorem 1}}, if the number of antennas in an array is smaller than the antenna-domain wideband boundary, the beam squint effect within this array can be ignored, and the channel can be approximated via {\bl\eqref{Eq: Section3_1_8}}. Therefore, the wideband boundaries provide a guidance for efficient subarray slicing, and then the wideband precoding vector can be designed to provide an improved SE of the wideband hybrid-field channel.

\begin{figure}
  \centering
  \includegraphics[scale=0.3]{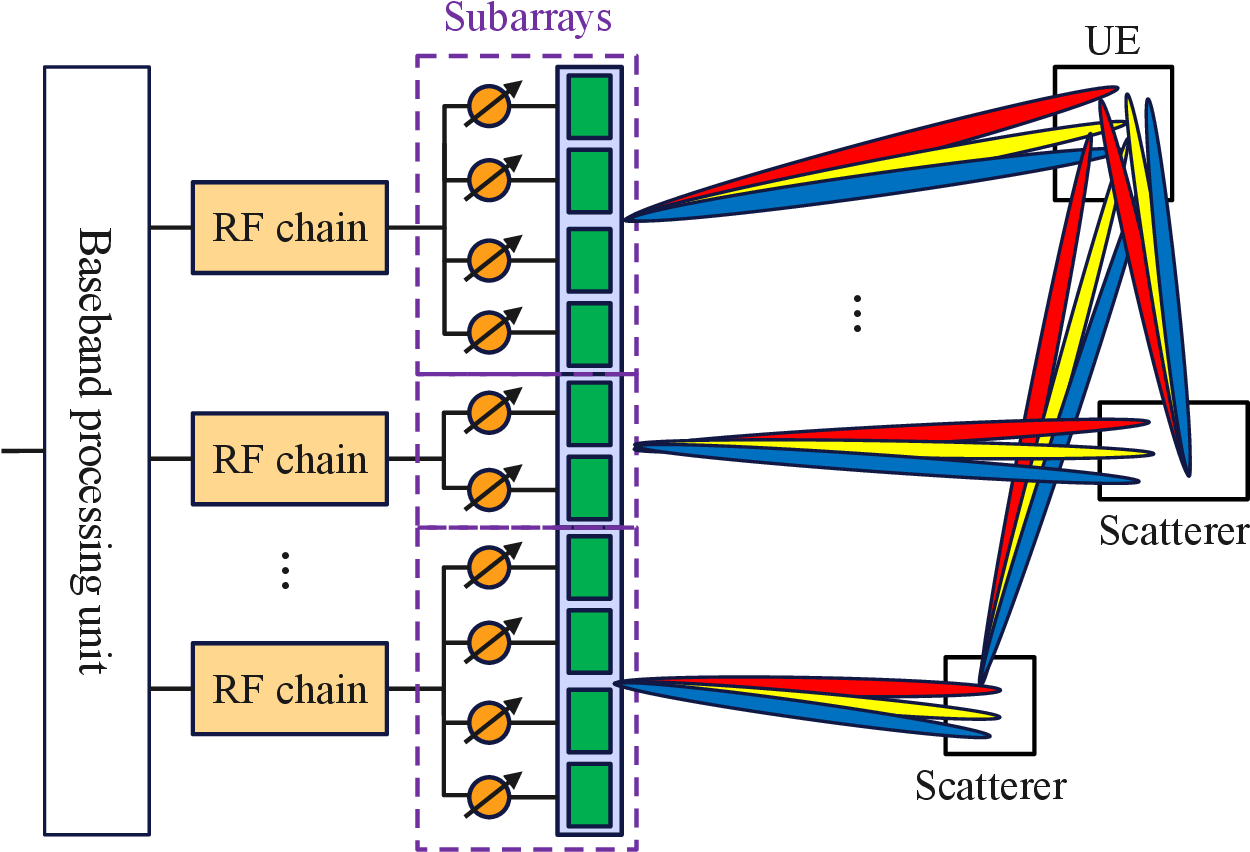}
  \caption{Antenna-domain channel slicing for wideband precoding.} \label{Fig: antenna-domain slicing}
\end{figure}

To reduce the hardware cost, subarray hybrid precoding architectures are generally considered in XL-MIMO systems \cite{hybrid_architecture2}. As shown in Fig.~\ref{Fig: antenna-domain slicing}, assume that the entire array is divided into $T$ subarrays, with the $t$-th subarray having $N_t$ antennas, and equipped with a single RF chain, for $t = 1, \ldots, T$, where $\sum \limits_{t=1}^T N_t = N$. When the BS transmits a downlink pilot to the UE, the received signal, ${\bf y}_{\rm wn} \in {\mathbb{C}^{M \times 1}}$, can be denoted as
\begin{equation}\label{Eq: Section4_1_2}
\left[{\bf{y}}_{\rm wn}\right]_{m} = \sqrt{P} \left[{\bf{H}}_{\rm wn}\right]_{:,m}^{\rm H} {\bf{F}}_{{\rm A},{\rm as}} {\bf f}_{{\rm D},m} x_m + z_m,
\end{equation}
for $m = 1,\ldots,M$, where $z_m$ is a complex additive white Gaussian noise (AWGN) with mean zero and variance $\sigma^2$; $x_m$ is the transmitted pilot at the $m$-th subcarrier, satisfying $\left|x_m\right| = 1$; ${\bf F}_{{\rm A},{\rm as}} \in {\mathbb{C}^{N \times T}} = {\rm blkdiag}\left({\bf f}_{{\rm as},1},\ldots,{\bf f}_{{\rm as},T}\right)$ denotes the frequency-independent analog precoding matrix, where ${\bf f}_{{\rm as},t} \in \mathbb{C}^{N_t \times 1}$ is the analog precoding vector of the $t$-th subarray, whose each element has constant amplitude, i.e., $\left| {{\left[ {\bf f}_{{\rm as},t} \right]}_n} \right| = 1$, for $n = 1,\ldots,N_t$, and $t = 1,\ldots,T$; ${\bf f}_{{\rm D},m} \in \mathbb{C}^{T \times 1}$ denotes the frequency-dependent digital precoding vector at the $m$-th subcarrier, satisfying the power constraint $\left\| {\bf F}_{{\rm A},{\rm as}} {\bf f}_{{\rm D},m} \right\| = 1$. Then, the downlink SE can be calculated as
\begin{equation}\label{Eq: Section4_1_3}
{\rm SE}_{\rm as} = \frac{1}{M} \sum\limits_{m = 1}^M {{{\log}_2} \left( {1 + \frac{P{{\left| {{{\bf f}_{{\rm D},m}^{\rm H}} {\bf F}_{{\rm A},{\rm as}}^{\rm H} {{\left[ {{\bf H}_{\rm wn}} \right]}_{:,m}}} \right|}^2}} {\sigma^2}} \right)}.
\end{equation}
The MRT digital precoding vector ${\bf f}_{{\rm D},m}$ can be designed as
\begin{equation}\label{Eq: Section4_1_4}
\begin{aligned}
{\bf f}_{{\rm D},m} = &\frac{{\bf F}_{{\rm A},{\rm as}}^{\rm H} \left[ {\bf H}_{\rm wn} \right]_{:,m}} {\left\| {\bf F}_{{\rm A},{\rm as}} {\bf F}_{{\rm A},{\rm as}}^{\rm H} \left[ {\bf H}_{\rm wn} \right]_{:,m} \right\|} \\= &\frac{{\bf F}_{{\rm A},{\rm as}}^{\rm H} \left[ {\bf H}_{\rm wn} \right]_{:,m}}{\sqrt{\sum\limits_{t = 1}^T N_t \left| {\bf f}_{{\rm as}, t}^{\rm H} \left[{\bf H}_{{\rm as},t}\right]_{:,m} \right|^2}},
\end{aligned}
\end{equation}
for $m = 1,\ldots,M$; ${{\bf{H}}_{{\rm{as}},t}} \in \mathbb{C}^{N_t \times M}$ denotes the channel of the $t$-th subarray, which can be expressed as
\begin{equation}\label{Eq: Section4_1_5}
\begin{aligned}
&{{\bf{H}}_{{\rm{as}},t}} = \sum_{l_{\rm F} = 1}^{L_{\rm F}} g_{l_{\rm F}} {\bf a}_{{\rm as}, t}\left(\theta_{l_{\rm F}} \right) {\bf p}^{\rm T} \left(\theta_{l_{\rm F}}, d_{l_{\rm F}} \right) \\+ &\sum_{l_{\rm N} = 1}^{L_{\rm N}} g_{l_{\rm N}} \left( {{{\bf{w}}_{{\rm{as}},t}}\left( {\theta_{l_{\rm N}} ,d_{l_{\rm N}}} \right){{\bf{p}}^{\rm{T}}}\left( {d_{l_{\rm N}},r_{l_{\rm N}}} \right)} \right) \odot {{\bf{Q}}_{{\rm{as}},t}}\left( {\theta_{l_{\rm N}} ,d_{l_{\rm N}}} \right),
\end{aligned}
\end{equation}
for $t = 1,\ldots,T$, where ${\bf Q}_{{\rm as},t}\left(\theta, d \right) \in \mathbb{C}^{N_t \times M}$ is the near-field beam squint matrix of the $t$-th subarray, satisfying ${\bf{Q}}_{\rm wn}\left( {\theta ,d} \right) = {\left[ {{\bf{Q}}_{{\rm{as}},1}^{\rm{T}}\left( {\theta ,d} \right), \cdots ,{\bf{Q}}_{{\rm{as}},T}^{\rm{T}}\left( {\theta ,d} \right)} \right]^{\rm{T}}}$, while ${\bf a}_{{\rm as},t}\left(\theta \right) \in \mathbb{C}^{N_t \times M}$ is the far-field antenna-domain steering vector of the $t$-th subarray, satisfying ${\bf{a}} \left( {\theta } \right) = {\left[ {{\bf{a}}_{{\rm{as}},1}^{\rm{T}}\left( {\theta} \right), \cdots ,{\bf{a}}_{{\rm{as}}, T}^{\rm{T}}\left( {\theta} \right)} \right]^{\rm{T}}}$. The near-field antenna-domain steering vector ${\bf w}_{{\rm as},t}\left(\theta,d\right) \in \mathbb{C}^{N_t \times 1}$ of the $t$-th subarray is given by
\begin{equation}\label{Eq: Section4_1_6}
{{\bf{w}}_{{\rm{as}},t}}\left( {\theta ,d} \right) = {\left[ {{e^{j\frac{{2\pi }}{c}{f_{\rm{c}}}\Delta {d_{{\rm{as}},t,1}}}}, \cdots ,{e^{j\frac{{2\pi }}{c}{f_{\rm{c}}}\Delta {d_{{\rm{as}},t,{N_t}}}}}} \right]^{\rm{T}}},
\end{equation}
where $\Delta {d_{{\rm{as}},t,n}} = {d_{{\rm{as}},t,n}} - {d_{{\rm{as}},t}}$; ${d_{{\rm{as}},t}}$ and ${d_{{\rm{as}},t,n}}$ represent the distance from the scatterer to the center and to the $n$-th antenna of the $t$-th subarray, respectively; ${d_{{\rm{as}},t,n}}$ can be defined as
\begin{equation}\label{Eq: Section4_1_7}
\begin{aligned}
&{d_{{\rm{as}},t,n}} =\\ &\sqrt {{d^2} - 2d\left( {N_t' + \delta_{N_t, n}} \right)s\theta  + {{\left( {N_t' + \delta_{N_t, n}} \right)}^2}{s^2}},
\end{aligned}
\end{equation}
for $n = 1,\ldots,N_t$, where $N_t' = - \frac{N}{2} + \sum_{u = 1}^{t - 1} N_u + \frac{N_t}{2}$, while ${d_{{\rm{as}},t}}$ is given by
\begin{equation}\label{Eq: Section4_1_8}
{d_{{\rm{as}},t}} = \sqrt {{d^2} - 2d {N_{t}'}s\theta  + \left(N_{t}' s\right)^2}.
\end{equation}
Then, the SE can be alternatively written as
\begin{equation}\label{Eq: Section4_1_9}
{\rm SE}_{\rm as} = \frac{1}{M} \sum\limits_{m = 1}^M {{{\log}_2} \left( {1 + \frac{P\left({\sum\limits_{t = 1}^T {\left| {{\bf f}_{{\rm as},t}^{\rm H}{{\left[ {{\bf H}_{{\rm as},t}} \right]}_{:,m}}} \right|^2}}\right)^2} {{\sum\limits_{t = 1}^T N_t \left| {\bf f}_{{\rm as}, t}^{\rm H} \left[{\bf H}_{{\rm as},t}\right]_{:,m} \right|^2} \sigma^2}} \right)}.
\end{equation}

To improve the SE of U6G XL-MIMO systems, the design of the analog precoding vector follows two principles: \textbf{\textit{1)}} Multipath components must be considered. Specifically, since the scale of the array should be limited to eliminate the effect of beam squint in each near-filed component, the entire array is sliced into subarrays, and each subarray focus on a single near-field component. \textbf{\textit{2)}} To enhance the spatial resolution, each near-field component in the subarray still attains the spherical wave features, thus maintaining a greater orthogonality between different paths (the orthogonality requires a relatively large scale of subarray). Specifically, the antenna-domain wideband and near-field boundaries of the $l_{\rm N}$-th near-field component are indicated by $\bar N_{{\rm wn}, l_{\rm N}}$ and $\tilde N_t$, respectively, for $l_{\rm N} = 1,\ldots,L_{\rm N}$, which have been introduced in Table~\ref{Tab: channel types}. Assume that all the near-field components have been sorted in descending order of power, i.e. $\left| g_1 \right| \ge \ldots \ge \left| g_{L_{\rm N}} \right|$. The $t$-th subarray focus on the $t'$-th near-field component, where $t' = t - \left\lfloor \frac{t - 1}{L_{\rm N}} \right\rfloor$. Consequently, the scale of the $t$-th subarray should satisfy $\tilde N_{t'} \le N_t \le \bar N_{{\rm wn}, {t'}}$, for $t = 1,\ldots,T$. Then, the beam squint effect of the $t'$-th near-field path in the $t$-th subarray can be ignored, thus ${\bf Q}_{{\rm as}, t} \left( \theta_{t'}, d_{t'} \right) \approx {\bf 1}_{N_t \times M}$. The $t$-th analog precoding vector (which considers the $t'$-th near-field path and all the far-field paths), ${\bf f}_{{\rm as},t} \in \mathbb{C}^{N_t \times 1}$, can then be designed as ${{\bf{f}}_{{\rm{as}},t}} = e^{j \angle {\cal B}_t}$, whereas ${\cal B}_t \in \mathbb{C}^{N_t \times 1}$ is given by
\begin{equation}\label{Eq: Section4_1_10}
{\cal B}_t = \sum\limits_{l_{\rm F} = 1}^{L_{\rm F}} g_{l_{\rm F}} {\bf a}_{{\rm as}, t} \left(\theta_{l_{\rm F}} \right) + g_{t'} {{\bf w}_{{\rm as},t}} \left( {\theta_{t'} ,d_{t'}} \right),
\end{equation}
where $t' = t - \left\lfloor \frac{t - 1}{L_{\rm N}} \right\rfloor L_{\rm N}$, for $t = 1,\ldots,T$. Note that in a real U6G XL-MIMO wireless environment, and due to the longer propagation distances, the power of the far-field components is much smaller than that of the near-field components. Thus, the effects of the far-field components can be ignored when designing the hybrid precoding vector. Moreover, due to the great spatial resolution when $N_{t}$ is large, $\left| {\bf w}_{{\rm as},t}^{\rm H} \left( \theta_i, d_i \right) {\bf w}_{{\rm as},t} \left( \theta_j, d_j \right) \right| \approx 0$ typically holds, for $1 \le i \ne j \le L_{\rm N}$. Then, the SE can be approximated as
\begin{equation}\label{Eq: Section4_1_11}
{\rm SE}_{\rm as} \approx \frac{1}{M} \sum\limits_{m = 1}^M \log_2 \left(1 + \frac{P \left(\sum\limits_{t = 1}^{T} \left| g_{t'} \right|^2 N_t^2 \right)^2} {\sum\limits_{t = 1}^T \left|g_{t'}\right|^2 N_t^3 \sigma^2} \right).
\end{equation}

\textbf{\textit{Discussion:}} A special case is considered, where $N_t = \frac{N}{T}$, for $t = 1,\ldots,T$. Then, the SE can be denoted as
\begin{equation}\label{Eq: Section4_1_12}
{\rm SE}_{\rm as} \approx \log_2 \\ \left( 1 + \frac{P N \sum\limits_{t = 1}^T \left| g_{t'} \right|^2}{T \sigma^2} \right).
\end{equation}
Note that when $ L_{N} = 1$, we have ${\rm SE}_{\rm wn} \approx \log_2 \left( {1 + \frac{PN\left| g_1\right|^2}{\sigma^2}}\right) = {\rm SE}_{\rm opt}$, where ${\rm SE}_{\rm opt}$ denotes the SE upper bound as in {\bl\eqref{Eq: Section3_3_6}}. Thus, the antenna-domain slicing scheme can efficiently mitigate the beam squint effect, and improve the SE for U6G XL-MIMO systems.

\subsection{Frequency-Domain Sub-Band Slicing}

\begin{figure}
  \centering
  \includegraphics[scale=0.26]{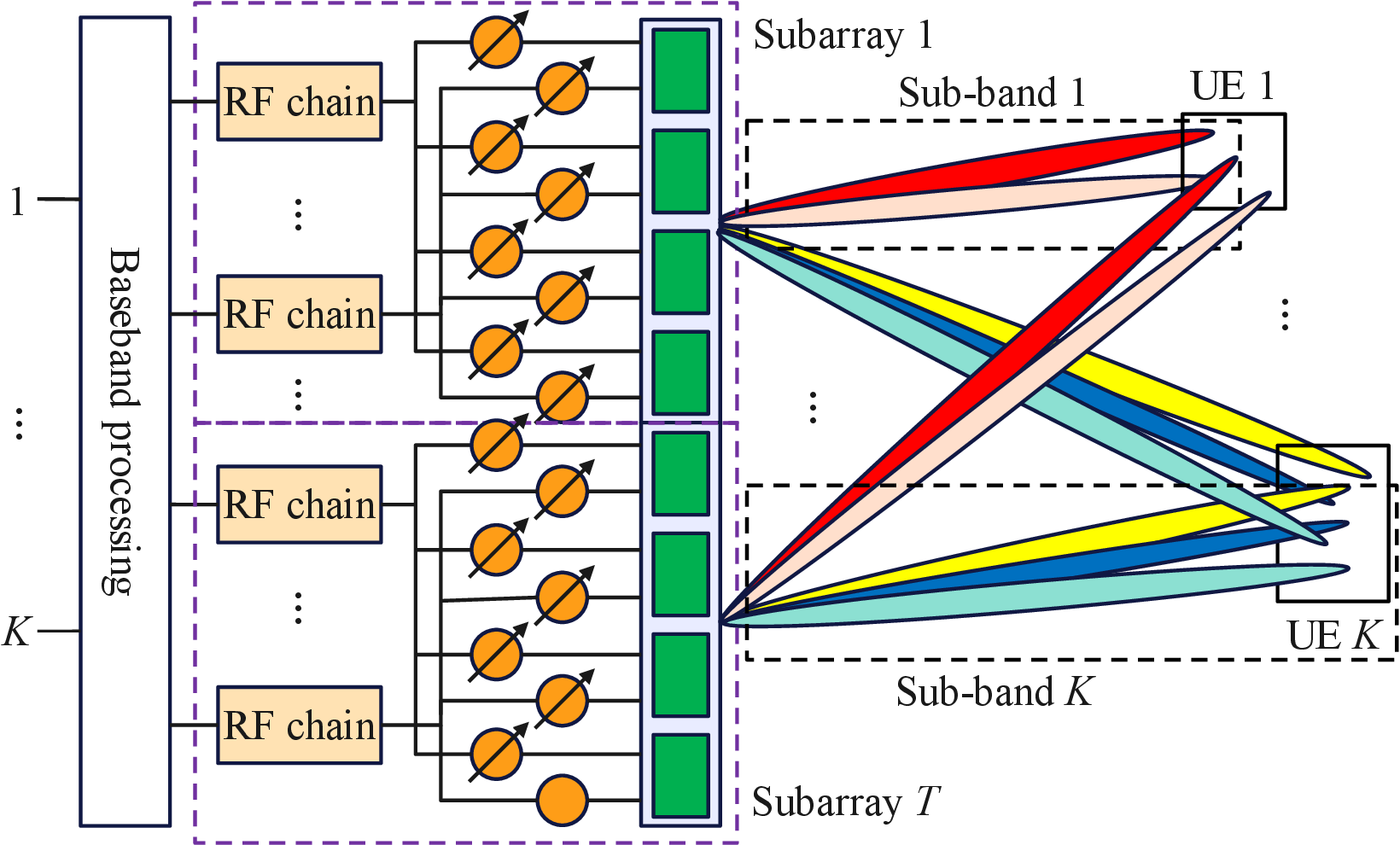}
  \caption{Frequency-domain channel slicing for wideband precoding.} \label{Fig: frequency-domain slicing}
\end{figure}

It is important to note that although slicing the entire array into subarrays can effectively improve the SE for U6G XL-MIMO systems under the beam squint effect, this approach requires modification of the scale of each subarray in accordance with real propagation communication. In authentic U6G XL-MIMO systems, the number and scale of subarrays are typically constant. Consequently, it is imperative to conduct a study on efficient wideband precoding in such a scenario. In 5G New Radio (NR) systems, different UEs are typically allocated distinct resource blocks (RBs) in the frequency domain. In the context of future U6G XL-MIMO systems, a pivotal challenge pertains to the evaluation of the bandwidth allocation to different UEs, with the objective of enhancing the SE under the beam squint effect. In addition to slicing the entire array into subarrays, the constraint of the bandwidth allocated to each UE has also been identified as a means of improving the SE, since the beam squint effect is caused by the large arrays and wide bandwidths simultaneously.

\begin{figure}
  \centering
  \includegraphics[scale=0.5]{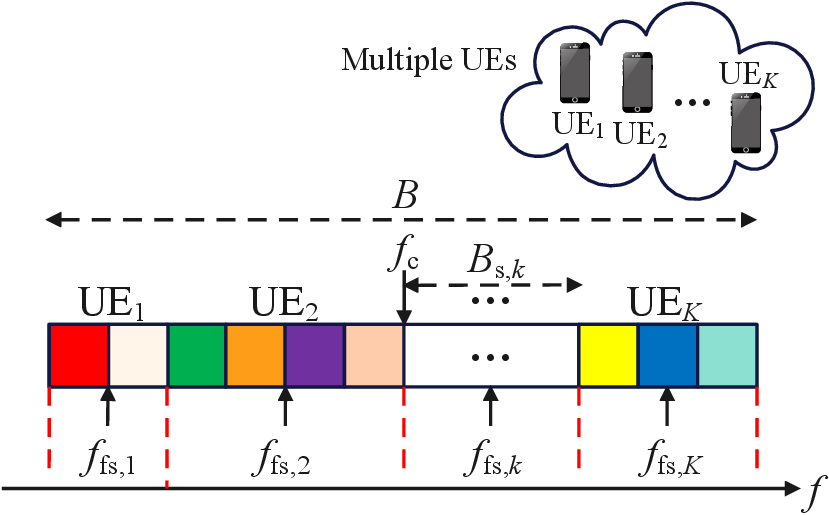}
  \caption{Sub-band slicing for multiple-UE resource allocation.} \label{Fig: resource allocation}
\end{figure}

The works in \cite{beamsquint6, lens_far, frequency_multiplexing2, frequency_multiplexing1, subband} have developed the concept of dividing the entire bandwidth into sub-bands to serve different UEs. However, additional switches and lenses are required in \cite{beamsquint6} and \cite{lens_far}, respectively, resulting in an increase in hardware cost. The approaches in \cite{frequency_multiplexing2, subband} serve each UE by only a single subarray, leading to a compromised SE due to the lower gain array. Consequently, a frequency-domain slicing scheme is proposed in this section, with the objective of enhancing the SE for U6G XL-MIMO systems with multipath channels. As shown in Fig.~\ref{Fig: frequency-domain slicing}, the entire array has been uniformly divided into $T$ subarrays, with each subarray having $N_{\rm s} = \frac{N}{T}$ antennas. Each RF chain is only connected to the antennas within a single subarray. To improve the SE for U6G XL-MIMO systems, we focus on slicing the entire bandwidth into sub-bands and allocating them to different UEs, with the objective of mitigating the beam squint effect. As shown in Fig.~\ref{Fig: resource allocation}, the number of UEs is indicated by $K$, and the bandwidth allocated for the $k$-th UE is denoted by $B_{{\rm s}, k}$, which contains $M_{{\rm s}, k} = \frac{B_{{\rm s}, k}}{\Delta f}$ subcarriers, for $k = 1,\ldots,K$. Specifically, the central frequency of the $k$-th sub-band is denoted as $\tilde f_{{\rm c},k}$, such that
\begin{equation}\label{Eq: Section4_2_1}
\tilde f_{{\rm c}, k} = f_{\rm c} - \frac{B}{2} + \sum\limits_{u = 1}^{k - 1} B_{{\rm s}, u} + \frac{B_{{\rm s}, k}}{2}.
\end{equation}
Note that the frequency of the $m_k$-th subcarrier in $B_{{\rm s}, k}$ is given by $\tilde f_{k, m_k} = \tilde f_{{\rm c}, k} + \delta_{M_{{\rm s}, k}, m_k} \Delta f$, where $\sum\limits_{k = 1}^K M_{{\rm s}, k} = M$, and $\sum\limits_{k = 1}^K B_{{\rm s}, k} = B$. The wideband channel of the $k$-th UE, ${\bf H}_{{\rm wn}, k} \in \mathbb{C}^{N \times M_{{\rm s}, k}}$, and the channel w.r.t. the $t$-th subarray, ${\bf H}_{{\rm as}, t, k} \in \mathbb{C}^{N_{\rm s} \times M_{{\rm s}, k}}$, can be denoted similarly to {\bl\eqref{Eq: Section4_1_1}} and {\bl\eqref{Eq: Section4_1_5}}, respectively, for $k = 1,\ldots,K$, and $t = 1,\ldots,T$. Therefore, the signal received by the $k$-th UE, ${\bf y}_{{\rm wn}, k} \in \mathbb{C}^{M_{{\rm s}, k}\times 1}$, can be expressed according to 
\begin{equation}\label{Eq: Section4_2_2}
\left[{\bf{y}}_{{\rm wn},k}\right]_{m_k} = \sqrt{P} \left[{\bf{H}}_{{\rm wn},k}\right]_{:,m_k}^{\rm H} {\bf{F}}_{{\rm A},{\rm fs},k} {\bf f}_{{\rm D},k,m_k} x_{k,m_k} + z_{k,m_k},
\end{equation}
for $m_k = 1,\ldots,M_{{\rm s}, k}$, where $z_{k,m_k}$ is a complex AWGN with mean zero and variance $\sigma^2$, for $k = 1,\ldots,K$; $x_{k,m_k}$ is the transmitted pilot at the $m_k$-th subcarrier to the $k$-th UE, satisfying $\left|x_{k,m_k}\right| = 1$; ${\bf F}_{{\rm A},{\rm fs},k} \in {\mathbb{C}^{N \times T}} = {\rm blkdiag}\left({\bf f}_{{\rm fs},1,k},\ldots,{\bf f}_{{\rm fs},T,k}\right)$ denotes the frequency-independent analog precoding matrix to the $k$-th UE, where ${\bf f}_{{\rm fs},t,k} \in \mathbb{C}^{N_{\rm s} \times 1}$ is the analog precoding vector of the $t$-th subarray, satisfying $\left| {{\left[ {\bf f}_{{\rm fs},t,k} \right]}_n} \right| = 1$, for $n = 1,\ldots,N_{\rm s}$, and $t = 1,\ldots,T$; ${\bf f}_{{\rm D},k,m_k} \in \mathbb{C}^{T \times 1}$ denotes the frequency-dependent digital precoding vector at the $m_k$-th subcarrier for the $k$-th UE, satisfying the power constraint $\left\| {\bf F}_{{\rm A},{\rm fs},k} {\bf f}_{{\rm D},k,m_k} \right\| = 1$. Then, the downlink average SE can be calculated as
\begin{equation}\label{Eq: Section4_2_3}
\begin{aligned}
{\rm SE}_{\rm fs} = &\frac{1}{K}\sum\limits_{k = 1}^K \frac{1}{M_{{\rm s},k}} \\ \times &\sum_{m_k = 1}^{M_{{\rm s},k}} \log_2 \left(1 + \frac{P \left| {\bf f}_{{\rm D},k,m_k}^{\rm H} {\bf F}_{{\rm A},{\rm fs},k}^{\rm H} \left[ {\bf H}_{{\rm wn},k} \right]_{m_k} \right|^2}{\sigma^2}\right).
\end{aligned}
\end{equation}

It is important to note that a frequency-domain wideband boundary has been provided to mitigate the beam squint effect. Assume that the frequency-domain wideband boundary of the $k$-th UE (in a single subarray and the $l_{{\rm N}}$-th near-field path) can be denoted as $\bar B_{{\rm wn},k,l_{\rm N}}$. Consequently, the bandwidth allocated for the $k$-th UE in the $t$-th subarray satisfies the constraint that $B_{{\rm s},k} \le \max\left\{\bar B_{{\rm wn},k,1},\ldots,\bar B_{{\rm wn},k,L_{{\rm N},k}}\right\}$, where $L_{{\rm N},k}$ denotes the number of near-field paths in ${\bf H}_{{\rm wn}, k}$. Subsequently, the beam squint effect in each subarray can be efficiently mitigated to improve the SE of each UE. Without loss of generality, assume that the number of far-field paths in ${\bf H}_{{\rm wn}, k}$ satisfies $L_{{\rm F}, k} = 0$, for $k = 1,\ldots,K$, due to the much lower received power due to the longer transmission distance. Note that when a frequency-independent MRT precoder is adopted, the normalized array gain of $\left[{\bf H}_{{\rm wn}, k}\right]_{m_k}$ can be denoted as
\begin{equation}\label{Eq: Section4_2_4}
\begin{aligned}
{\eta}_{k,m_k} = &\frac{\left|\sum\limits_{l_{\rm N} = 1}^{L_{{\rm N},k}} e^{j\frac{2\pi}{c}\delta_{{M_{{\rm s},k}},{m_k}} \Delta f\left(r_{k,l_{{\rm N},k}} + d_{k,l_{{\rm N},k}}\right) } \left|g_{k,l_{\rm N}}\right|^2\right|} {\sqrt{\sum\limits_{l_{\rm N} = 1}^{L_{{\rm N},k}} \left|g_{k,l_{\rm N}}\right|^2}} \\= &\frac{\left|\sum\limits_{l_{\rm N} = 1}^{L_{{\rm N},k}} e^{j \Delta \omega_{k,m_k,l_{{\rm N},k}}} \left|g_{k,l_{\rm N}}\right|^2\right|} {\sqrt{\sum\limits_{l_{\rm N} = 1}^{L_{{\rm N},k}} \left|g_{k,l_{\rm N}}\right|^2}},
\end{aligned}
\end{equation}
where $\omega_{k,m_k,l_{{\rm N},k}} = \frac{2\pi}{c}\delta_{{M_{{\rm s},k}},{m_k}} \Delta f\left(r_{k,l_{{\rm N},k}} + d_{k,l_{{\rm N},k}} - \tilde D_{k} \right)$, and $\tilde D_{k} = \frac{1}{L_{{\rm N},k}} \sum\limits_{l_{{\rm N}, k}}^{L_{{\rm N},k}} \left(r_{k,l_{{\rm N}, k}} + d_{k,l_{{\rm N}, k}}\right)$. To mitigate the effect of $\omega_{k,m_k,l_{{\rm N},k}}$ between different paths in the SE, the following constraint has to be fulfilled $\mathop {\max }\limits_{m_k,l_{{\rm N},k}} \left| \omega_{k,m_k,l_{{\rm N},k}}\right| \le \kappa_{\rm f}\pi$. Thus, a bandwidth constraint can be obtained as $B_{{\rm s}, k} \le \tilde B_{k}$, where
\begin{equation}\label{Eq: Section4_2_5}
\tilde B_k = \frac{k_{\rm f} c}{\mathop {\max }\limits_{l_{{\rm N},k}} \left| r_{k,l_{{\rm N},k}} + d_{k,l_{{\rm N},k}} - \tilde D_{k} \right| }. 
\end{equation}
Consequently, the bandwidth allocated to the $k$-th UE should satisfy $B_{{\rm s}, k} \le \min \left\{ \bar B_{{\rm  wn}, k, 1}, \ldots, \bar B_{{\rm  wn}, k, L_{{\rm  N}, k}}, \tilde B_k \right\}$. As shown in Fig.~\ref{Fig: frequency-domain slicing}, the MRT digital precoding vector ${\bf f}_{{\rm D},k,m_k}$ is indicated by {\bl\eqref{Eq: Section4_1_4}}, while the frequency-independent analog precoder can be designed as ${\bf f}_{{\rm fs}, t, k} = e^{j \angle {\cal C}_{t, k}}$. ${\cal C}_{t, k} \in \mathbb{C}^{N_{\rm s} \times 1}$ is given by
\begin{equation}\label{Eq: Section4_2_6}
\begin{aligned}
{\cal C}_{t,k} = &\sum\limits_{l_{{\rm N},k} = 1}^{L_{{\rm N},k}} g_{k,l_{{\rm N},k}} e^{j\frac{2\pi}{c}\left(\tilde f_{{\rm c},k} - f_{\rm c}\right) \left(r_{k,l_{{\rm N},k}} + d_{k,l_{{\rm N},k}}\right)} \\ \times &{\bf w}_{{\rm as}, t} \left(\theta_{k,l_{{\rm N},k}}, d_{k,l_{{\rm N},k}} \right).
\end{aligned}
\end{equation}
Subsequently, the average SE can be denoted as 
\begin{equation}\label{Eq: Section4_2_7}
\begin{aligned}
{\rm SE}_{\rm fs} \approx &\frac{1}{K}\sum\limits_{k = 1}^K \frac{1}{M_{{\rm s},k}} \sum_{m_k = 1}^{M_{{\rm s},k}} \\ & \log_2  \left(1 + \frac{P \left( \sum\limits_{t = 1}^T \left(  \sum\limits_{n = 1}^{N_{\rm s}} \left| \left[{\bf H}_{{\rm as},t,k}\right]_{m_k} \right| \right)^2 \right)^2}{ \sum\limits_{t = 1}^T N_{\rm s} \left(  \sum\limits_{n = 1}^{N_{\rm s}} \left| \left[{\bf H}_{{\rm as},t,k}\right]_{m_k} \right| \right)^2 \sigma^2} \right).
\end{aligned}
\end{equation}

\textbf{\textit{Discussion:}} A special case is considered, where $L_{{\rm N},k} = 1$, and $\left|g_{k, 1}\right| = \left|g\right|$, for $k = 1,\ldots,K$. Then, the average SE can be expressed as
\begin{equation}\label{Eq: Section4_2_8}
{\rm SE}_{\rm fs} \approx \log_2 \\ \left( 1 + \frac{P N \sum\limits_{t = 1}^T \left| g \right|^2}{\sigma^2} \right) = {\rm SE}_{\rm opt},
\end{equation}
which indicates that the frequency-domain slicing scheme is also an efficient method to greatly improve the SE for U6G wideband XL-MIMO systems.

\section{Simulation Results}\label{Sec: Simulation Results}

\begin{figure}
  \centering
  \includegraphics[scale=0.32]{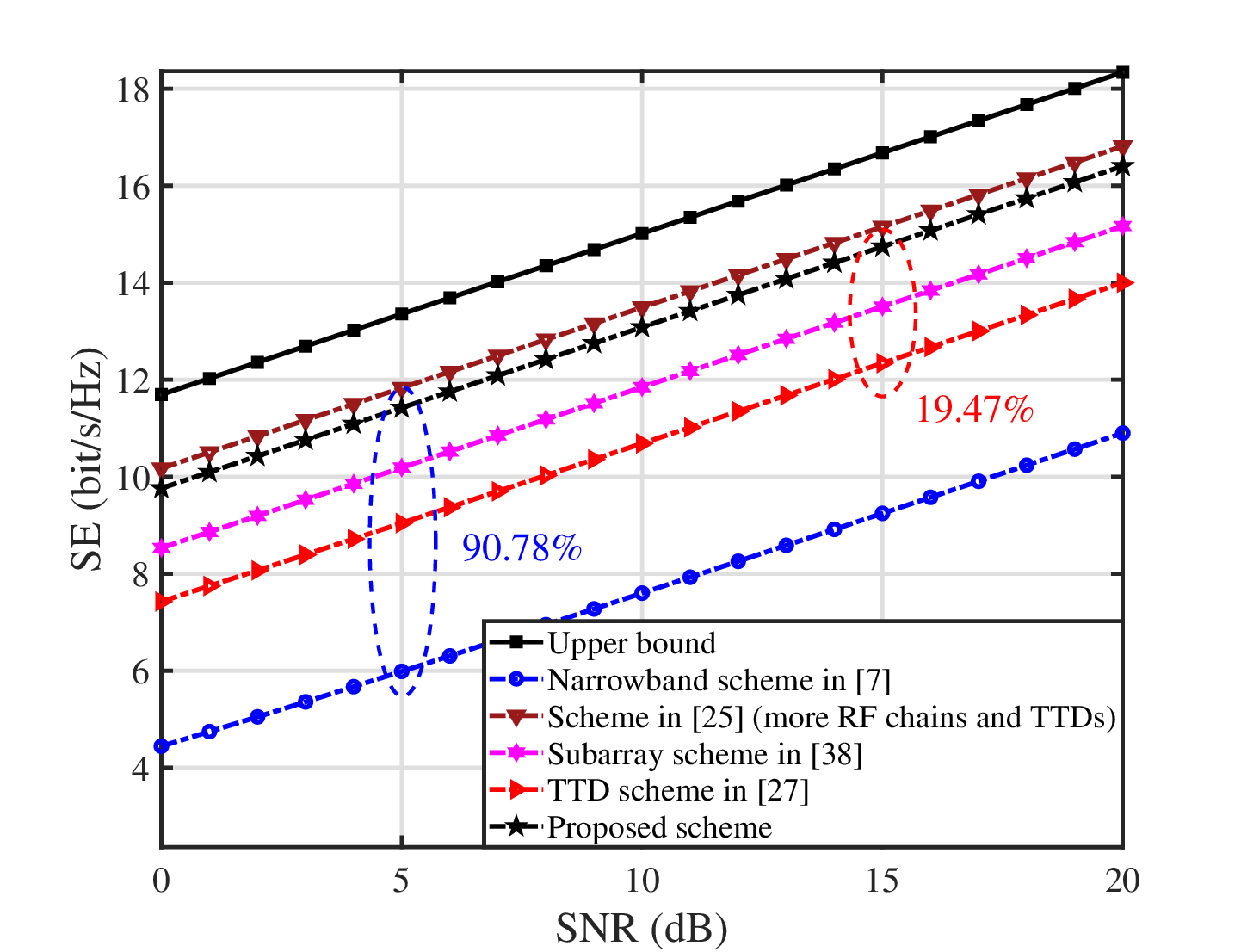}
  \caption{SE against the SNR (antenna-domain slicing).} \label{Fig: simulation1}
\end{figure}

In this section, the SE of the proposed wideband boundaries and the channel slicing scheme in U6G XL-MIMO systems is evaluated. The system parameters are set as follows: $N = 1024$; $f_{\rm c} = 7$ GHz; $B = 600$ MHz, and $M = 256$. Moreover, $L_{\rm N} = 4$; $L_{\rm F} = 1$; $\theta$ and $\left\{d,r\right\}$ are generated uniformly within $\left(-1,1\right)$ and $\left[10,100\right]$ m, respectively. The number of subarrays in other schemes is set to 32, and the SNR is 10 dB.

\begin{figure}
  \centering
  \includegraphics[scale=0.32]{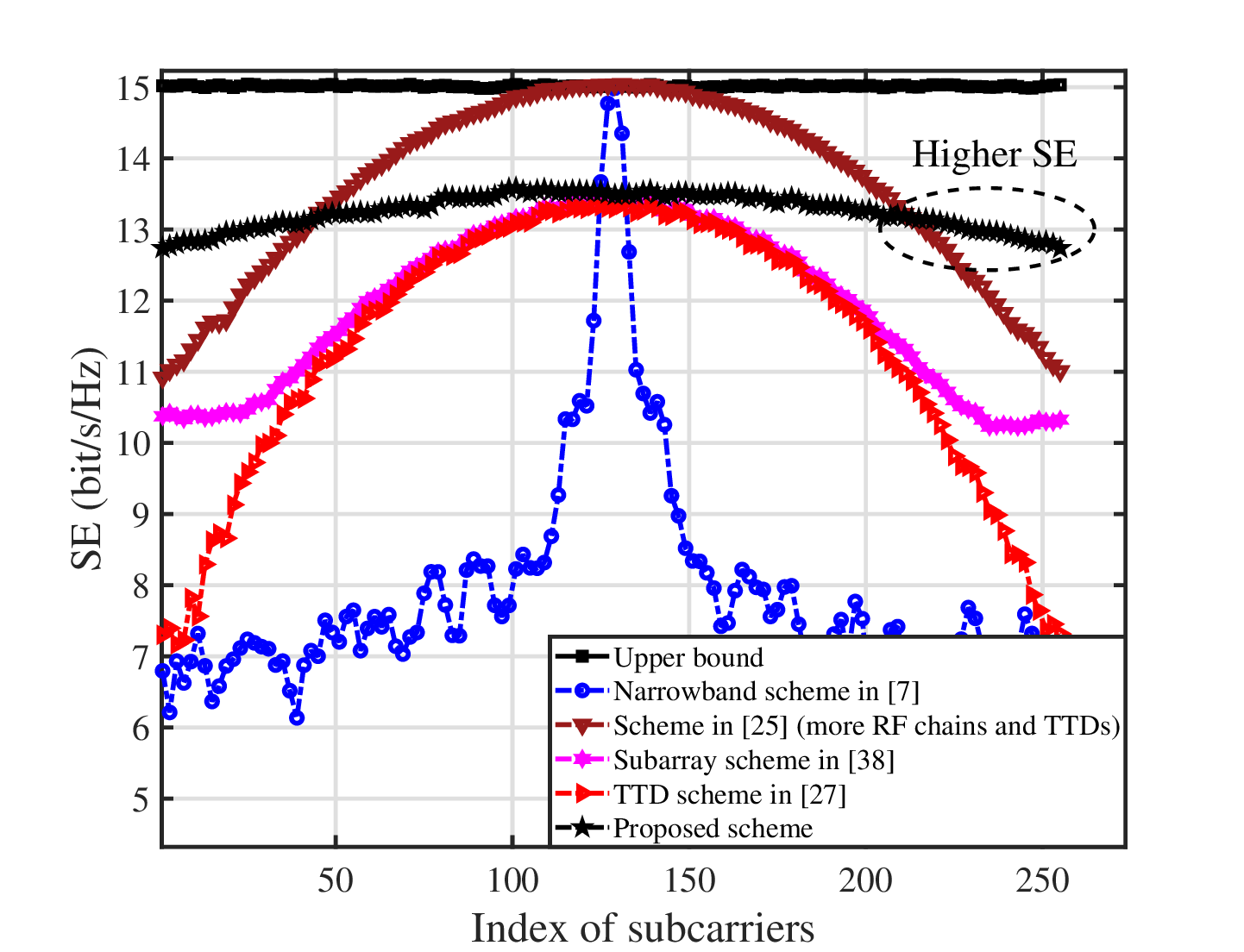}
  \caption{SE against the index of the subcarriers (antenna-domain slicing).} \label{Fig: simulation2}
\end{figure}

The SE of U6G XL-MIMO systems with antenna-domain slicing is evaluated firstly. As shown in Fig.~\ref{Fig: simulation1} and Fig.~\ref{Fig: simulation2}, the beam squint effect will significantly compromise the SE of the conventional narrowband hybrid precoding scheme in \cite{SE3}. Although the works in \cite{multiple_beam, frequency_multiplexing2} focus on mitigating the beam squint effect by dividing the subarrays, the effect of hybrid-field multiple paths in U6G XL-MIMO channels have not been considered. Compared to other schemes, the proposed antenna-domain slicing scheme can significantly enhance the SE by reasonably designing the scale of each subarray based on the wideband boundaries, and considering the presence of multiple paths in U6G XL-MIMO systems. As shown in Fig.~\ref{Fig: simulation1}, the proposed scheme can achieve an improvement in SE of $90.78\%$ and $19.47\%$ than the narrowband scheme in \cite{SE3} and the TTD scheme in \cite{multiple_beam}, respectively. Moreover, as shown in Fig.~\ref{Fig: simulation2}, the SE has been improved across the entire bandwidth when $B$ is a constant, in particular for the subcarriers that are far from the central carrier frequency.

\begin{figure}
  \centering
  \includegraphics[scale=0.32]{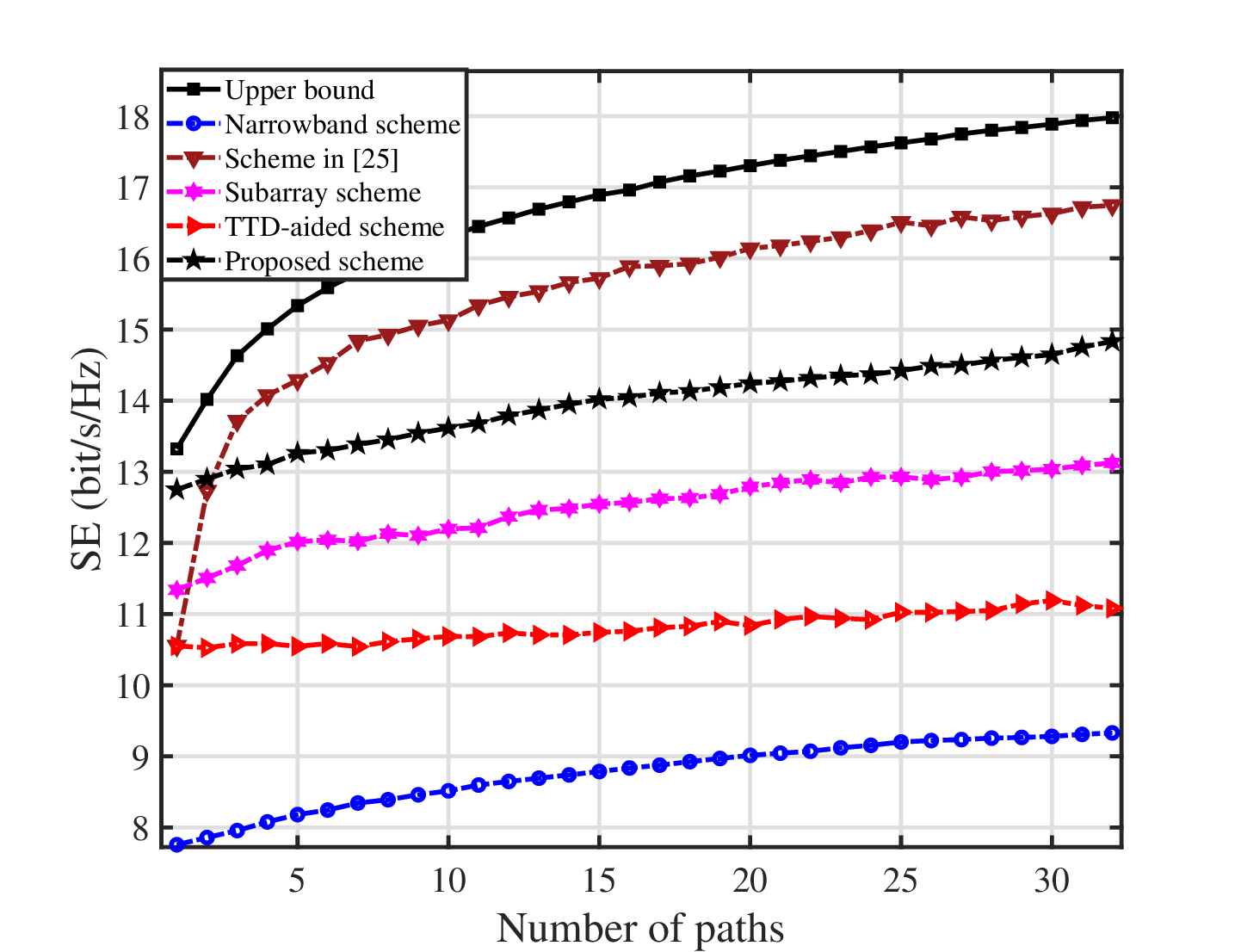}
  \caption{SE against the number of near-field paths (antenna-domain slicing).} \label{Fig: simulation3}
\end{figure}

The TTD scheme in \cite{multiple_beam} mitigates the beam squint effect by deploying $T$ TTD lines, while the proposed scheme can achieve a higher SE without TTD lines, by slicing the subarrays based on the proposed wideband boundaries and considering the presence of multiple paths. Although the work in \cite{TTD_far3} has a better SE than the proposed scheme, it requires $\left(L_{\rm N} + L_{\rm F}\right)$ times the number of RF chains and TTD lines than that of the scheme in \cite{multiple_beam} (i.e., a total of $\left(L_{\rm N} + L_{\rm F}\right)T$ TTD lines). Note that the proposed scheme can still achieve an SE close to that of the scheme in \cite{TTD_far3}. Furthermore, the SE against the number of near-field paths has also been evaluated in Fig.~\ref{Fig: simulation3}. Since multiple paths in U6G XL-MIMO channels have been considered, the SE of the antenna-domain slicing scheme improves with an increase in the number of paths. When only a single path exists, the proposed scheme can achieve an SE close to the upper bound.

\begin{figure}
  \centering
  \includegraphics[scale=0.32]{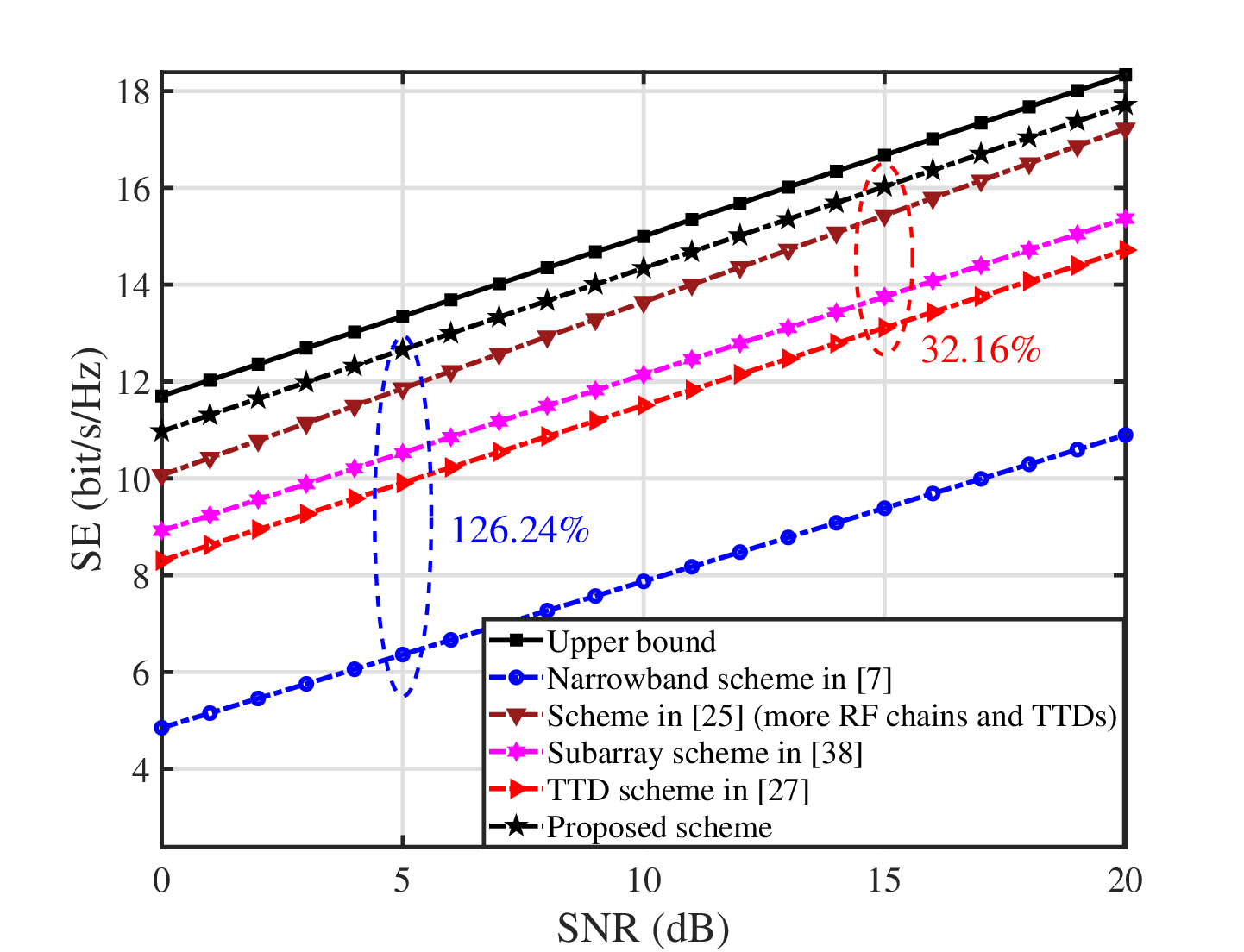}
  \caption{SE against the SNR (frequency-domain slicing).} \label{Fig: simulation4}
\end{figure}

\begin{figure}
  \centering
  \includegraphics[scale=0.32]{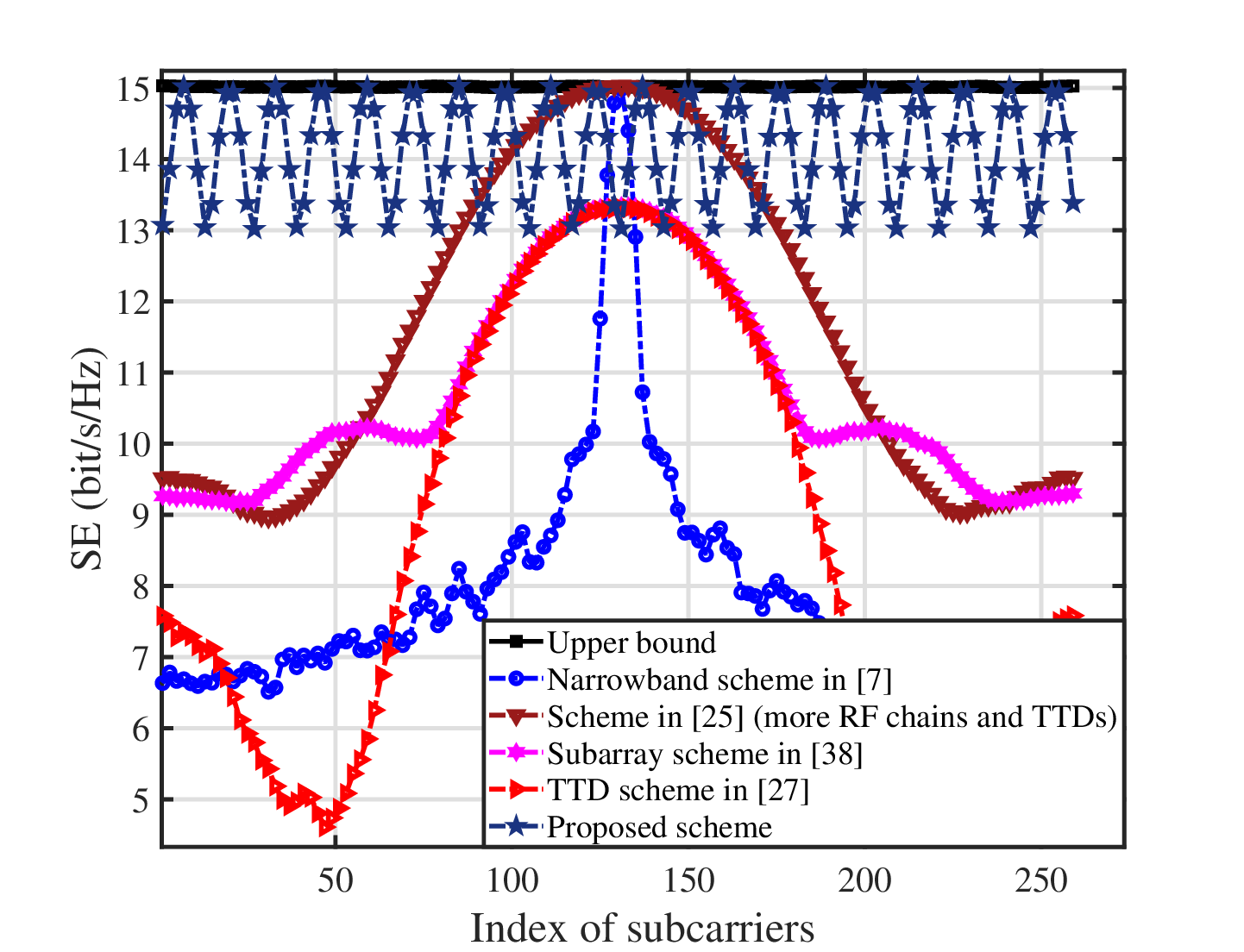}
  \caption{SE against the index of the subcarriers (frequency-domain slicing).} \label{Fig: simulation5}
\end{figure}

The SE of U6G XL-MIMO systems with frequency-domain slicing is then evaluated, where $T = 8$. As shown in Fig.~\ref{Fig: simulation4} and Fig.~\ref{Fig: simulation5}, the SE of the proposed scheme against the SNR and $B$ is evaluated. As shown in Fig.~\ref{Fig: simulation4}, the proposed scheme achieves a higher SE than other schemes. Specifically, the proposed scheme has an improvement in SE of $126.24\%$ and $32.16\%$ than the narrowband scheme in \cite{SE3} and the TTD scheme in \cite{multiple_beam}, respectively. Note that the proposed frequency-domain slicing method has an enhanced SE than the antenna-domain slicing method, since the entire subarray has been divided into $T$ subarrays. As shown in Fig.~\ref{Fig: simulation5}, the proposed frequency-domain slicing scheme can improve the SE of U6G XL-MIMO systems across the entire bandwidth.

\begin{figure}
  \centering
  \includegraphics[scale=0.32]{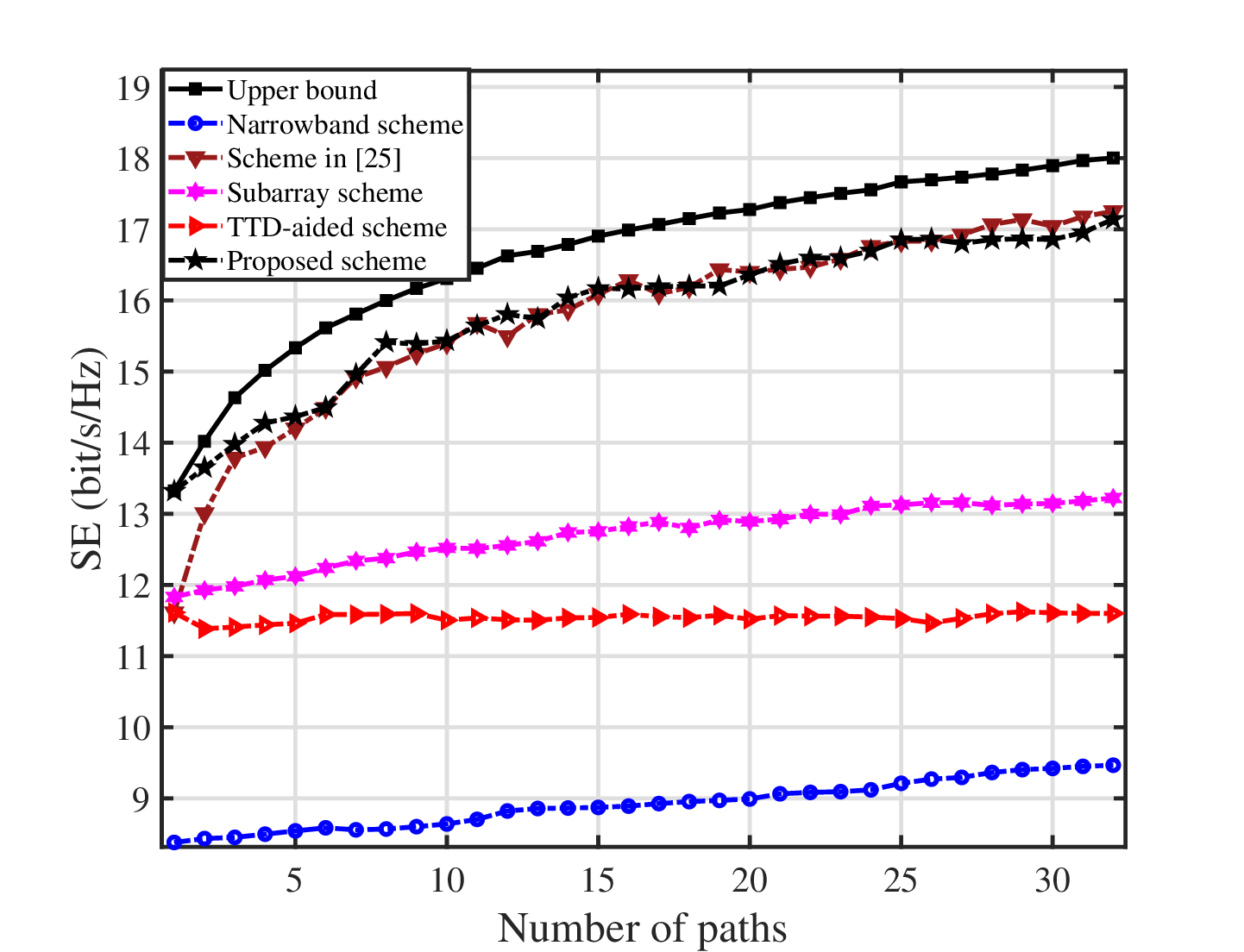}
  \caption{SE against the number of near-field paths (frequency-domain slicing).} \label{Fig: simulation6}
\end{figure}

\begin{figure}
  \centering
  \includegraphics[scale=0.32]{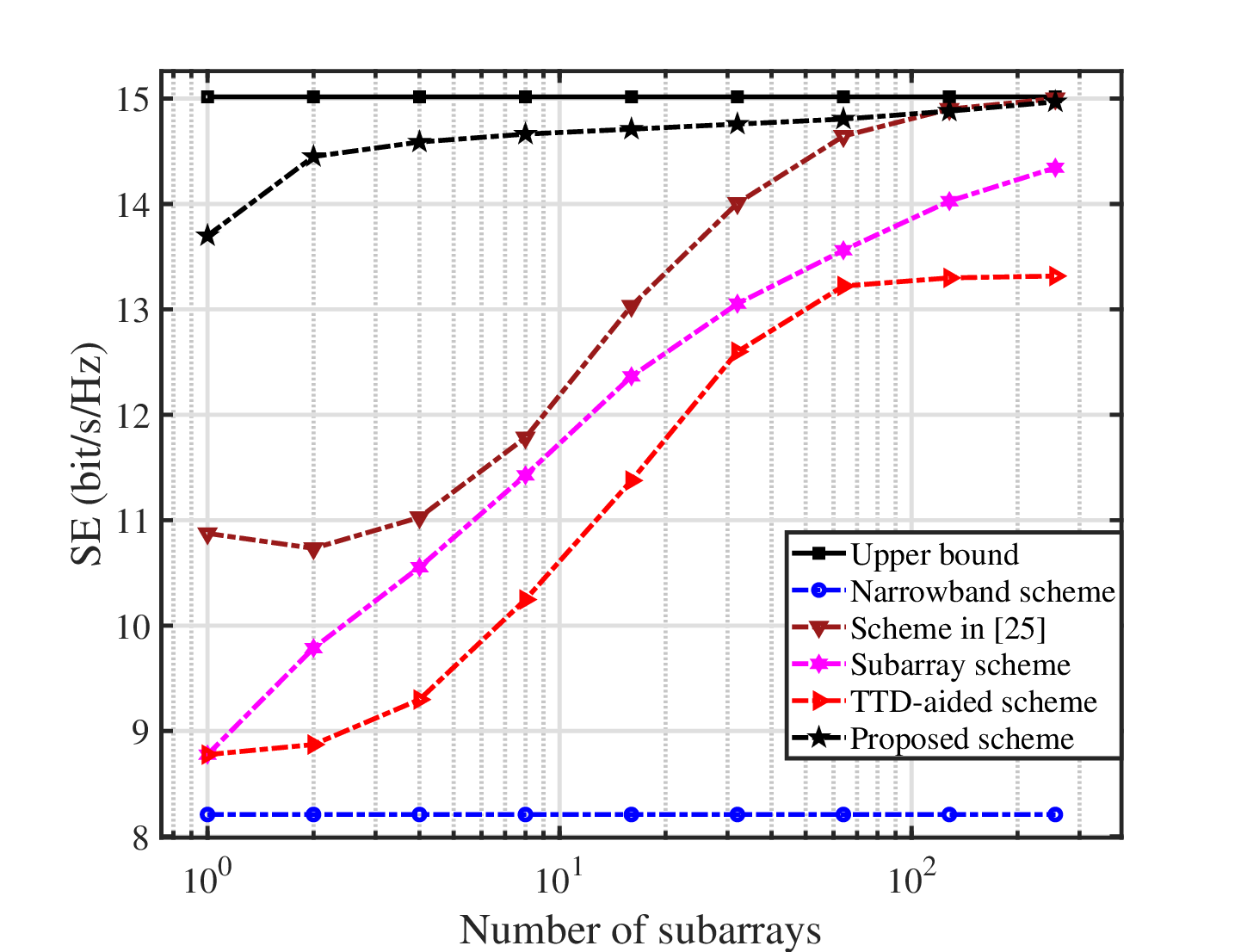}
  \caption{SE against the number of subarrays (frequency-domain slicing).} \label{Fig: simulation7}
\end{figure}

The SE against the number of near-field paths and subarrays is also evaluated in Fig.~\ref{Fig: simulation6} and Fig.~\ref{Fig: simulation7}, respectively. As shown in Fig.~\ref{Fig: simulation6}, the SE of the proposed frequency-domain slicing scheme improves with an increase in the number of near-field paths, since the effect of multiple-path channels has been considered, while it achieves an SE close to the upper bound in the single-path scenario. Moreover, the proposed frequency-domain slicing scheme achieves a higher SE than other schemes, since the effects of beam squint and phase variations between different paths at different subcarriers have been mitigated. As shown in Fig.~\ref{Fig: simulation7}, the proposed scheme still has a high SE when the value of $T$ is small. This is because the beam squint effect can be mitigated from the antenna-domain and frequency-domain perspectives simultaneously.

\section{Conclusion}\label{Sec: Conclusion}

In this paper, a U6G XL-MIMO system with a hybrid precoding architecture was considered. To infer whether the beam squint issue persists in U6G XL-MIMO systems, precise frequency-domain and antenna-domain wideband boundaries were derived from the near-field and far-field perspectives, respectively. This can inform the system designer whether the beam squint effect should be considered in a specific system setting. Moreover, a fascinating phenomenon was observed. As the array size and the bandwidth increase, the near-field effect invariably precedes the beam squint effect. To improve the SE when the beam squint effect is non-negligible, a channel slicing scheme was introduced to mitigate the beam squint effect, while an efficient wideband precoding scheme was designed from the antenna-domain and frequency-domain perspectives, respectively. Since the characteristics of U6G XL-MIMO channels and architectures have been considered, the proposed wideband precoding scheme can enhance the SE across the entire bandwidth. The simulation results demonstrated that the beam squint issue remains pronounced in the near-field region of the U6G XL-MIMO systems. Furthermore, the proposed channel slicing scheme based on wideband boundaries has the potential to significantly improve the SE by designing appropriate wideband precoding.

In future work, the joint use of subarray slicing and sub-band slicing methods will be studied, with a comprehensive consideration of overhead and performance. In addition, the exploration will also extend to wideband precoding schemes for time-varying U6G XL-MIMO systems.

\begin{appendices}

\section{Proof of Theorem 1}

It is evident that the phase shift in ${\bf Q}_{\rm wn}\left(\theta_l, d_l \right)$ is related to the extremely large-scale array via {\bl\eqref{Eq: Section2_6}} and the wide bandwidth via {\bl\eqref{Eq: Section2_9}} simultaneously. Consequently, two terms $\kappa_{\rm a} \pi$ and $\kappa_{\rm f} \pi$ are adopted to indicate the phase shift thresholds from the antenna-domain and frequency-domain perspectives, respectively. Specifically, the beam squint effect of a path can be ignored when the maximum phase shifts in the antenna domain and frequency domain are less than $\kappa_{\rm a} \pi$ and $\kappa_{\rm f} \pi$, respectively, which can be denoted as
\begin{equation}\label{Eq: Section3_1_6}
\mathop {\max }\limits_{m,n} \angle {{\bf{Q}}_{{\rm{wn}}}}\left( {\theta_l, d_l} \right) < \left(\kappa_{\rm a} + \kappa_{\rm f}\right) \pi.
\end{equation}
Substituting {\bl\eqref{Eq: Section3_1_5}} into {\bl\eqref{Eq: Section3_1_6}}, we can get
\begin{equation}\label{Eq: SectionA_1}
B < \frac{\left(\kappa_{\rm a} + \kappa_{\rm f} \right) c}{f_{\rm wn}\left(\theta_l,d_l\right)},
\end{equation}
and thus the frequency-domain wideband boundary can be obtained. For the antenna-domain wideband boundary, due to the symmetry of the incident signals from two sides of the array normal direction (i.e., the maximum distance variation $f_{\rm wn}\left(\theta,d\right)$ is an even function w.r.t. $\theta$, i.e., $f_{\rm wn}\left(\theta_l,d_l\right) = f_{\rm wn}\left(-\theta_l,d_l\right)$, for $-1 < \theta_l < 0$), only the scenario $0 \le \theta_l < 1$ is analyzed in the following sections. Note that when $-1 < \theta_l < 0$, the same conclusion can be reached. When $0 \le \theta_l < 1$, {\bl\eqref{Eq: Section3_1_4}} can also be expressed as
\begin{equation}\label{Eq: Section3_1_11}
{f}_{\rm wn}\left(\theta_l,d_l\right) = \sqrt{{\cal A}_1\left(N-1\right)^2+{\cal A}_2\left(N-1\right)+d_l^2}-d_l,
\end{equation}
where ${\cal A}_1 = \frac{\lambda_{\rm c}^2}{16}$, ${\cal A}_2 = \frac{\lambda_{\rm c}d_l\left|\theta_l\right|}{2}$. Substituting {\bl\eqref{Eq: Section3_1_11}} into {\bl\eqref{Eq: Section3_1_5}} and {\bl\eqref{Eq: Section3_1_6}} respectively, we can get
\begin{equation}\label{Eq: SectionA_2}
\sqrt{{\cal A}_1\left(N-1\right)^2 + {\cal A}_2 \left(N-1\right)+d_l^2} < \frac{\left(\kappa_{\rm a} + \kappa_{\rm f} \right) c}{B} + d_l,
\end{equation}
and then
\begin{equation}\label{Eq: Section3_1_12}
\begin{aligned}
{\cal F} \left(N-1\right) = &{\cal A}_1\left(N-1\right)^2 + {\cal A}_2\left(N-1\right) - {\cal A}_3 \\= &{\cal A}_1 \left(\left(N-1\right) + \frac{{\cal A}_2}{2{\cal A}_1}\right)^2 -\frac{{\cal A}_2^2+4{\cal A}_1{\cal A}_3}{4{\cal A}_1} < 0,
\end{aligned}
\end{equation}
where
\begin{equation}\label{Eq: SectionA_3}
{\cal A}_3 = \frac{\left( \kappa_{\rm a} + \kappa_{\rm f} \right)^2 c^2+2\left( \kappa_{\rm a} + \kappa_{\rm f} \right)cd_lB}{B^2}.
\end{equation}
Note that ${\cal F} \left(N-1\right)$ is a quadratic convex function of one variable with respect to $N-1$, and the minimum point is located at $N-1 = -\frac{{\cal A}_2}{2{\cal A}_1} < 0$, because ${\cal A}_1 > 0$, and ${\cal A}_2 > 0$. Thus, for $N \ge 1$, ${\cal F} \left(N-1\right)$ is a monotonically increasing function w.r.t. $N$. Then, based on {\bl\eqref{Eq: Section3_1_12}} and $N \ge 1$, we have
\begin{equation}\label{Eq: SectionA_4}
0 \le {N -1 } < \frac{-{\cal A}_2+\sqrt{{\cal A}_2^2+4{\cal A}_1{\cal A}_3}}{2{\cal A}_1}.
\end{equation}
Therefore, \textbf{\textit{Theorem 1}} is proved. \qed

\section{Proof of Theorem 2}

In contrast to a near-field path, the maximum distance variation in the far-field perspective can be simplified as
\begin{equation}\label{Eq: Section3_1_21}
\begin{aligned}
&\mathop {\max }\limits_n \left| {\Delta {d_{n,l}}} \right| \\ \approx &\mathop {\max }\limits_n\left| { - {\delta _{N,n}}s\theta_l } \right| = \frac{{\left( {N - 1} \right){\lambda _{\rm{c}}}\left|\theta_l\right| }}{4} = f_{\rm wf}\left(\theta_l \right),
\end{aligned}
\end{equation}
while the maximum phase shift in ${\bf Q}_{\rm wf}\left(\theta_l\right)$ can be calculated as
\begin{equation}\label{Eq: SectionB_1}
\mathop {\max }\limits_{m,n} \angle {{\bf{Q}}_{{\rm{wf}}}}\left(\theta_l \right) = \frac{\pi B}{c}f_{\rm wf}\left(\theta_l\right).
\end{equation}
To eliminate the beam squint effect, the maximum phase shift in ${\bf Q}_{\rm wf} \left(\theta_l\right)$ should satisfy
\begin{equation}\label{Eq: SectionB_2}
\mathop {\max }\limits_{m,n} \angle {{\bf{Q}}_{{\rm{wf}}}}\left( {\theta_l} \right) < \left(\kappa_{\rm a} + \kappa_{\rm f}\right) \pi.
\end{equation}
Substituting {\bl\eqref{Eq: Section3_1_21}} into {\bl\eqref{Eq: SectionB_2}}, we can get
\begin{equation}\label{Eq: SectionB_3}
B < \frac{\left( \kappa_{\rm a} + \kappa_{\rm f} \right) c}{f_{\rm wf} \left(\theta_l\right)},
\end{equation}
and
\begin{equation}\label{Eq: SectionB_4}
N < \frac{4 \left( \kappa_{\rm a} + \kappa_{\rm f} \right) f_{\rm c}}{B\left|\theta_l\right|} + 1.
\end{equation}
Therefore, \textbf{\textit{Theorem 2}} is proved. \qed

\section{Proof of Theorem 3}

It is important to note that the boundaries between the far-field and near-field regions can be calculated by
\begin{equation}\label{Eq: SectionC_1}
\mathop {\max }\limits_n \frac{2\pi}{c}f_{\rm c}\Delta d_{n,l} = \mathop {\max }\limits_n \frac{2\pi}{c}f_{\rm c} f_{\rm wn} \left(\theta, d\right) < {\kappa}_{\rm a} \pi,
\end{equation}
and thus the near-field boundaries can be denoted as
\begin{equation}\label{Eq: Section3_1_25}
\tilde N = \frac{-{\cal A}_2 + \sqrt{{\cal A}_2^2 + 4 {\cal A}_1 {\cal A}_5}}{2 {\cal A}_1} + 1,
\end{equation}
when $B$ is constant. If $N < \tilde N$, the path can be modeled as a far-field path with planar wave model, while when $N \ge \tilde N$, the path must be modeled as a near-field path with spherical wave model. Note that
\begin{equation}\label{Eq: SectionC_2}
\begin{aligned}
{\cal A}_5 = &\frac{k_{\rm a}^2c^2 + 4k_{\rm a}cd_lf_{\rm c}}{4f_{\rm c}^2} \\= &\frac{k_{\rm a}^2c^2}{4f_{\rm c}^2} + \frac{k_{\rm a}cd_l}{f_{\rm c}} < \frac{\left(k_{\rm a} + k_{\rm f}\right)^2c^2}{4f_{\rm c}^2} + \frac{2\left(k_{\rm a} + k_{\rm f} \right) cd_l}{f_{\rm c}} \\ < & \frac{\left(k_{\rm a} + k_{\rm f}\right)^2c^2}{B^2} + \frac{2\left(k_{\rm a} + k_{\rm f} \right) cd_l}{B} \\ = & \frac{\left( \kappa_{\rm a} + \kappa_{\rm f} \right)^2 c^2 + 2\left( \kappa_{\rm a} + \kappa_{\rm f} \right)cd_lB}{B^2} = {\cal A}_3,
\end{aligned}
\end{equation}
since $k_{\rm f} > 0$ and $f_{\rm c} > B$. It is evident that $\tilde N < \bar N_{\rm wn}$ always holds, since ${\cal A}_5 < {\cal A}_3$. Therefore, it can be concluded that $N < \tilde N$ and $N \ge \bar N_{\rm wn}$ are mutually exclusive from a near-field perspective, which is the judgment conditions for a WF path model (that cannot hold simultaneously).

In addition, even from a far-field perspective, we can also get $\tilde N < \bar N_{\rm wf}$, when $B$ is constant. Substituting {\bl \eqref{Eq: Section3_1_21}} into {\bl \eqref{Eq: SectionC_1}}, $\tilde N$ can be approximately expressed as
\begin{equation}\label{Eq: SectionC_3}
\begin{aligned}
\tilde N \approx & \frac{2 k_{\rm a}}{\left|\theta_l\right|} + 1 < \frac{4\left(k_{\rm a} + k_{\rm f}\right)}{\left|\theta_l\right|} + 1 \\ = & \frac{4\left(k_{\rm a} + k_{\rm f}\right) B}{B\left|\theta_l\right|} + 1 < \frac{4\left(k_{\rm a} + k_{\rm f}\right) f_{\rm c}}{B\left|\theta_l\right|} + 1 = \bar N_{\rm wf},
\end{aligned}
\end{equation}
since $k_{\rm f} > 0$ and $f_{\rm c} > B$. Therefore, $\tilde N < \min \left\{ \bar N_{\rm wn}, \bar N_{\rm wf} \right\}$ always holds, and the judgment conditions for a WF path model that $N < \tilde N$ and $N \ge \bar N_{\rm wn}$ are always mutually exclusive from both near-field and far-field perspectives.

Furthermore, when $N$ is constant and according to {\eqref{Eq: SectionC_1}}, if an UE is within the far-field region of the BS, the carrier frequency must satisfy
\begin{equation}\label{Eq: SectionC_4}
f_{\rm c} < \frac{\kappa_{\rm a} c}{2 f_{\rm wn} \left(\theta, d\right)} < \frac{\left(\kappa_{\rm a} + \kappa_{\rm f} \right)c}{f_{\rm wn} \left(\theta, d\right)} = {\bar B}_{\rm wn} < {\bar B}_{\rm wf},
\end{equation}
where $k_{\rm f} > 0$. Note that $B < \min \left\{{\bar B}_{\rm wn}, {\bar B}_{\rm wf}\right\}$ always holds as $B$ increases, since the bandwidth must satisfy $B < f_{\rm c}$. It can be concluded that the far-field criterion {\bl \eqref{Eq: SectionC_4}} and the wideband criterion $B \ge \bar B_{\rm wn}$ (or $B \ge \bar B_{\rm wf}$) are mutually exclusive. Hence, \textbf{\textit{Theorem 3}} is proved. \qed

\end{appendices}

\end{document}